\documentclass[11pt]{article}

\usepackage{geometry,booktabs}  
\usepackage[T1]{fontenc}  
\usepackage[utf8]{inputenc}  
\usepackage{lineno}  
\usepackage{fancyhdr}  
\usepackage{enumitem}  
\usepackage{hyperref}  
\usepackage{titling}  
\usepackage{natbib}  
\usepackage{mathtools}  
\usepackage{titlesec}  
\usepackage{lastpage}  
\usepackage{comment}
\usepackage[english]{babel}  
\usepackage{amsmath}  
\usepackage{amsfonts}  
\usepackage{amssymb}  
\usepackage{wasysym}  
\usepackage{bbm}  
\usepackage{array}  
\usepackage{xr}  
\usepackage{verbatim} 
\usepackage{float} 
\usepackage{algorithm}
\usepackage{graphicx,subfigure}
\usepackage{amsthm}
\usepackage[dvipsnames]{xcolor}

\usepackage[noend]{algpseudocode}
\usepackage{natbib}

\setcitestyle{round,numbers}
\def\bp{{\mathbf p}}

\newcommand{\email}[1]{\href{mailto:{#1}}{{#1}}}

\newcommand{\keywords}[1]{\textbf{Keywords}: {#1}}

\newcommand{\optincludegraphics}[2][]{}

\newcommand{\optinput}[1]{}

\newtagform{brackets}{[}{]}
\usetagform{brackets}

\usepackage[capitalise,noabbrev]{cleveref}
\title{Nonparametric Modeling of Diffusion MRI Signal in Q-space}

\begin{document}

{\noindent\LARGE\bf \thetitle}

\bigskip

\begin{flushleft}\large
	Arkaprava Roy\textsuperscript{1},
	Zhou Lan\textsuperscript{2},
	Zhengwu Zhang\textsuperscript{3}
\end{flushleft}

\bigskip

\noindent
\begin{enumerate}[label=\textbf{\arabic*}]
\item Department of Biostatistics, University of Florida, Gainesville, FL, USA.
\item Brigham and Women's Hospital, Harvard Medical School, Boston, MA, USA.
\item Department of Statistics and Operations Research, University of North Carolina at Chapel Hill, Chapel Hill, NC, USA.
\end{enumerate}

\bigskip


\textbf{*} Corresponding to:

\indent\indent
\begin{tabular}{>{\bfseries}rl}
Arkaprava Roy		& {\email{arkaprava.roy@ufl.edu}}	\\		
Zhou Lan & {\email{zlan@bwh.harvard.edu}}\\
Zhengwu Zhang & {\email{zhengwu\_zhang@unc.edu}} \\
\end{tabular}

\vfill





\begin{abstract}
This paper describes a novel nonparametric model for modeling diffusion MRI signals in q-space.
In q-space, diffusion MRI signal is measured for a sequence of magnetic strengths (b-values) and magnetic gradient directions (b-vectors). 
We propose a Poly-RBF model, which employs a bidirectional framework with polynomial bases to model the signal along the b-value direction and Gaussian radial bases across the b-vectors. The model can accommodate sparse data on b-values and moderately dense data on b-vectors.
The utility of Poly-RBF is inspected for two applications: 1) prediction of the dMRI signal, and 2) harmonization of dMRI data collected under different acquisition protocols with different scanners. Our results indicate that
the proposed Poly-RBF model can more accurately predict the unmeasured diffusion signal than its competitors such as the Gaussian process model in {\tt Eddy} of FSL. Applying it to harmonizing the diffusion signal can significantly improve the reproducibility of derived white matter microstructure measures. 
\end{abstract}

\bigskip

\pagebreak

\section{Introduction}
Diffusion-weighted magnetic resonance imaging (dMRI) is a technique that measures the movement of water molecules to examine tissue micro-structure \citep{stejskal}. In imaging of the brain, dMRI measures the water diffusion in a region of interest (e.g., each voxel) containing many water molecules. The measured dMRI signal is generated from the average of all spins of water molecules in the voxel. The ensemble average of spins in the voxel (also called diffusion propagator or EAP) is denoted as $P_d({\boldsymbol r})$, where $\boldsymbol r$ represents the relative displacement in some diffusion time $t$. Under the narrow pulse setting \cite{mitra1995effects}, the EAP is related to diffusion signal $S({\boldsymbol q})$ (normalized by $b=0$ image) through the Fourier relationship \cite{karger1983propagator,tuch_2004}: $P_d({\boldsymbol r}) = \mathcal{F}(S({\boldsymbol q}))$, where $\mathcal{F}$ denotes the Fourier transform, and  ${\boldsymbol q}$ is diffusion wavevector decided by sequencing protocol ($\boldsymbol{q} = {\gamma \delta \boldsymbol{G}}/{2 \pi}$, which is a function of the gyromagnetic ratio $\gamma$, pulse duration $\delta$, and the gradient vector $\boldsymbol{G}$). While the EAP provides important information about the water diffusivity, to better understand WM fiber structure, researchers often derive a function called orientation distribution function (ODF), defined as $\psi(\boldsymbol u) = C \int P_d(r \boldsymbol u) dr $, where $\boldsymbol u$ is a unit direction, and $C$ here is a constant to make $\psi$ a probability density function (PDF). 
After deconvolution of $\psi(\boldsymbol u)$, a shaper version of ODF called fiber ODF (fODF) \cite{descoteaux2008deterministic} has been widely used for tractography and structural connectome analysis. 

\begin{figure}
    \centering
    \includegraphics[width = 0.8\textwidth]{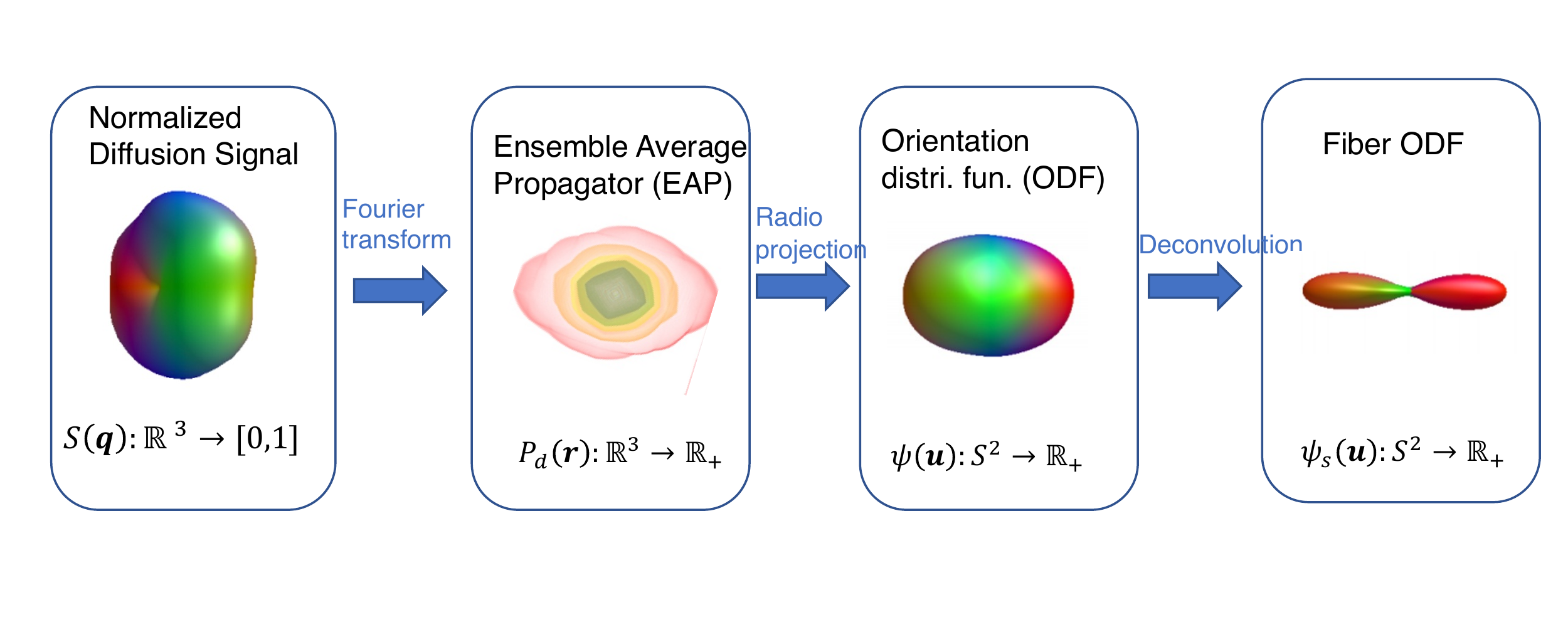}
    \caption{An illustration of the pipeline of computing fODF from normalized diffusion signal. The transformations from the normalized diffusion signal to the fODF are all linear.}
    \label{fig:odfmodel}
\end{figure}

Figure \ref{fig:odfmodel} presents common transformations from diffusion signal $S(\boldsymbol q)$ to fODF which is usually made in analyzing dMRI. Interestingly, all these transformations are linear, highlighting the importance of getting a good measure or estimation of $S(\boldsymbol q)$. For example, in Q-ball Imaging (QBI) \citep{tuch_2003, tuch_2004}, one can directly estimate an ODF function through the Funk–Radon transform (FRT) of $S(\boldsymbol q)$. In \cite{descoteaux_2007}, the authors showed that the FRT under spherical harmonics has a linear and analytical form. Moreover, the most popular constrained spherical deconvolution (CSD) algorithm \cite{tournier2007robust} relies on the linear relationship between $S(\boldsymbol q)$ and fODF. In real applications,  dMRI signal is measured in a domain called q-space, with parameters controlled by $\boldsymbol{q}$ defining the b-value and b-vector. Now denoting the measured signal as $S_v(b,{\bf p})$ along b-vector ${\bf p}$ ($\bf p \in \mathbb{S}^2$) at the $b$ b-value for voxel $v$, 
the dMRI data are mostly collected over an extremely sparse grid of b-values (mostly 2 to 4 b-values) and a moderately dense grid of b-vectors. In this paper, we are interested in the problem of accurately estimating $S_v$ from the observed sparse dMRI data.

There is a decent amount of literature on estimating diffusion signal function $S_v$ motivated by different applications. Anderson et al. \cite{andersson2015non} proposed a Gaussian process (GP) model to estimate the mean diffusion signal  $S_v(b,{\bf p})$ conditional on observed data. Most importantly, they implemented the GP model in the FSL {\tt eddy} toolbox \cite{andersson2016incorporating} for outlier frame detection and replacement. Therefore, it is one of the most used models for estimating $S_v$ for a given b-value and b-vector. Ning and colleagues \cite{ning2015estimating} proposed to use the radial basis function (RBF) to model $S_v$, and Cheng et al. \cite{cheng2010model} proposed to use Spherical Polar Fourier Expression to estimate the diffusion signal function. Instead of focusing on accurately estimating $S_v$, the formulations in \cite{ning2015estimating,cheng2010model} emphasize the simplicity of getting an estimate of EAP from their estimates of $S_v$. There are also various existing methods to estimate $S_v(b,\cdot): \mathbb{S}^2 \rightarrow \mathbb{R}_+$ for one given $b$ value, e.g., \cite{tuch_2004,descoteaux_2007,consagra2022optimized} among many others.

In this paper, we aim to accurately estimate $S_v(b,{\bf p})$ from noisy observed dMRI data collected at a highly sparse grid of b-values and a moderately dense grid of b-vectors. We formulate it into a nonparametric functional data estimation problem, relying on the following key observations on the structure of the diffusion signal: 1) given a fixed $\bp$, $\log(S_v(\cdot,{\bf p}))$ is a simple smooth function of $b$  which can be well modeled through polynomials \cite{tang2019diffusion}; and 2) given a fixed $b$, $\log(S_v(b,\cdot))$ is an unrestricted functional data on $\mathbb{S}^2$, which can be modeled through a linear combination of radial basis functions (RBF). 

\begin{figure}[htbp]
    \centering
    \begin{tabular}{cc}
        \includegraphics[width=6cm]{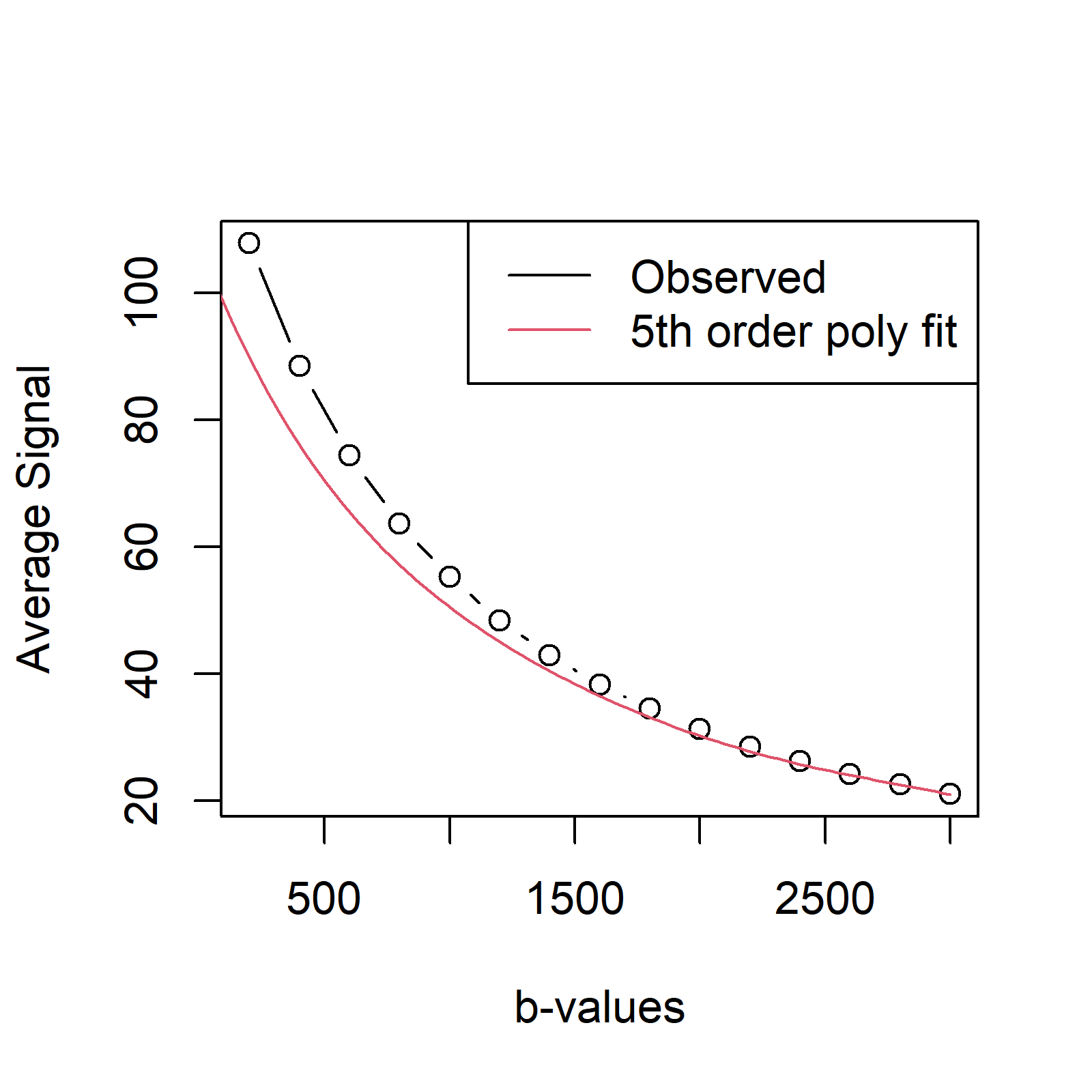} & 
        \includegraphics[width=6cm]{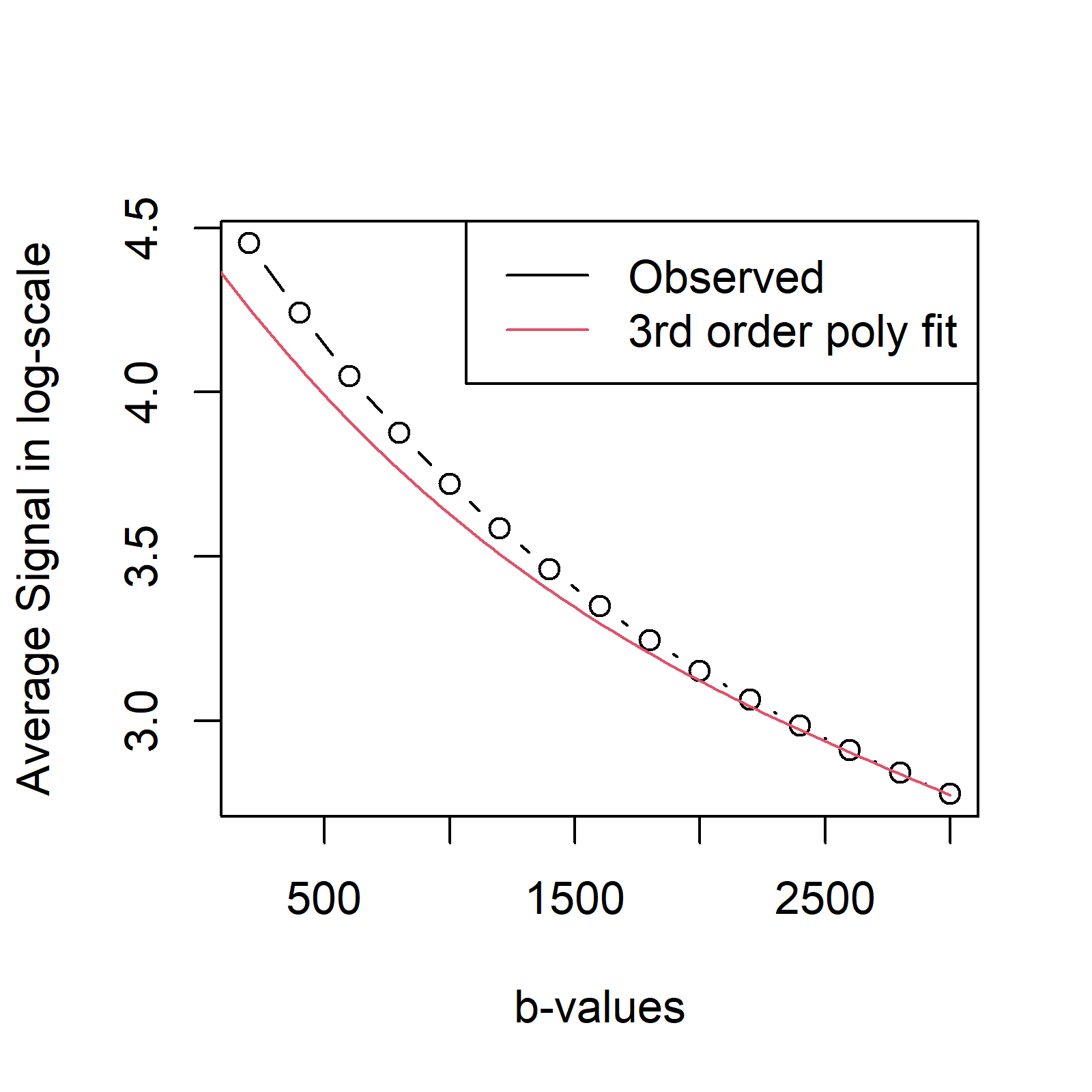} \\
    \end{tabular}
    \caption{Average raw signal (left) and average log-transformed signals (right) are plotted for different b-values using the CFIN dataset from the \texttt{dipy} package.}
    \label{fig:signalvsb}
\end{figure}

Most existing approaches model the diffusion signal $S_v$ directly \cite{cheng2010model,ning2015estimating,andersson2015non}. Although this direct estimation provides some convenience in calculating several diffusivity properties (see \cite{ning2015estimating}), constraint optimization routines were considered to ensure the fitted $S_v \geq 0$, making the implementation computationally expensive. Our proposal of modeling $S_v$ in the log-signal has two main advantages: 1) it eliminates the signal positive constraint, and 2) the log-transform helps to reduce model complexity, as shown in Figure~\ref{fig:signalvsb} where we plot $S_v(b)$ and $\log(S_v(b))$ across a grid of $b$ values (CFIN data with 15 unique non-zero b-values from  {\tt dipy} \citep{garyfallidis2014dipy} was used to create the plot). 
Here $\log(S_v)$ exhibits reduced non-linearity than $S_v$.
We further fit this data using polynomial bases in $b$. The cross-validated optimal order of polynomial for $\log(S_v)$ turns out to be 3 and for $S_v$ it is 5. This illustration also motivates us to use polynomial bases while modeling the variability of $\log(S_v(b,{\bf p}))$ with $b$. 

The utility of RBF in modeling diffusion signal is established in \cite{ning2015estimating}.
However, their model training required much denser grids on the q-space. Since the bandwidth-matrix for the RBF bases was kept fixed, outer shells would automatically require more bases than the inner shells to cover the space. In comparison, we take a bidirectional modeling approach, modeling the $b$-direction using polynomial bases and parallelly applying RBF bases for the $\bp$-direction. This modeling choice helps us significantly reduce the number of bases needed to approximate the diffusion signal.

We illustrate two important applications of our model. In the first one, we show that our model can be potentially used for better outlier frame detection and motion estimation in dMRI since it has superior performance in predicting diffusion signal over the main competitors in \cite{ning2015estimating,andersson2015non}. In the second one, we show the application of our method on dMRI data harmonization. Harmonization has become an important pre-processing step since existing neuroimaging studies often collected dMRI under different acquisition protocols with different scanners. In particular, different b-values and b-vectors can be used in different studies, producing batch effects when combining the data sets for a joint analysis. The traditional way of handling the batch effect is to use mixed-effect linear models such as ComBat \cite{johnson2007adjusting,fortin2017harmonization} to harmonize the diffusion-derived metrics, such as fractional anisotropy (FA) and some of the neurite orientation and dispersion density imaging (NODDI) metrics. However, the impacts of scanning protocols and scanners on diffusion-derived metrics are hardly linear, making the ComBat method ineffective in such data harmonization. Instead, we use our method to predict diffusion signals on a fixed grid of b-values and b-vectors to harmonize the scanning protocols first and then use ComBat to remove the scanner effect. 
We demonstrate the advantage of using the proposed hybrid technique for data harmonization in the experiment section.

The organization of the rest of the paper is as follows. In the next section, we illustrate our proposed model along with its implementation
and details on the two main applications of our proposed method using two data sets, and Section~\ref{sec:exp} presents our experimental results on real data. 
Finally, we conclude with some discussions and possible extensions in Section~\ref{sec:discussion}. 
The additional figures are given in the supplementary materials. 

\section{Methods}
\label{sec:method}
The dMRI data is usually collected over an extremely sparse grid of b-values and a moderately dense grid of gradient directions, the b-vectors. Let $(b_m,\bp_m)$ be the (b-value, b-vector) pair for $m$-$th$ dMRI sample. For most dMRI data, we have about 2 to 4 unique b-values and about 30 to 100 b-vectors per b-value. The signal is expected to vary smoothly with b-values and b-vectors, which is the key assumption of our proposed approach.

\subsection{Statistical modeling of dMRI signal}
Denote the measured diffusion signal (normalized to the b=0 image) as $S_v(b_m,\bp_m)$ at the voxel $v$ at some $\{(b_m,\bp_m)$ for $m=1,...,M$.
Note that there may be only 3 or 4 unique values in $\{b_m\}_{1\leq m\leq M}$.
Our proposed functional model for the dMRI data collected at $v$-$th$ voxel is, 
\begin{equation} \label{eq1}
\begin{split}
    \log\Big(S_{v}(b_m,\bp_m)\Big)=&f_{v}(b_m,\bp_m)+e_{v_m,\bp_m}\\
    f_v(b_m,\bp_m)=&\sum_{k=1}^K\theta_{v,k}(\bp_m)b_m^k,\\
    \theta_{v,k}(\bp_m)=&\sum_{\ell=1}^{2N} \beta_{v,k,\ell} \exp\left(-\frac{\|\bp_m-\hat{\bp}_{\ell}\|_2^2}{2h^2}\right)I\{\|\hat{\bp}_{\ell}-\bp_m\|_2<3h\},
\end{split}
\end{equation}
where we consider the tapered Gaussian kernels with bandwidth $h$ and $\{\hat{\bp}_1,\ldots,\hat{\bp}_{2N}\}$ are the centers such that $\hat{\bp}_i=-\hat{\bp}_{i+N}$ and correspondingly $\beta_{v,k,i}=\beta_{v,k,i+N}$ to ensure that the function $\theta_{v,k}(\bp_m)$ is symmetric.
Since, the b-vectors $\bp_{m}$'s are supported on $\mathbb{S}^2$, we set the $\hat{\bp}_{\ell}$ as unit-vectors and further, ensure that for all $1\leq i<j\leq N$, we have $-1<\hat{\bp}_i^T\hat{\bp}_j<1$ such that all the directions are unique.
We refer our proposed method as Poly-RBF$(K,N)$.

In our case, the centers $\hat{\bp}_{\ell}$'s and design points $\bp_m$'s are all supported on a sphere. Thus, we have $\|\hat{\bp}_{\ell}-\bp_m\|_2^2=2(1-\hat{\bp}_{\ell}^T\bp_m)=2\{1-\cos(\alpha_{\hat{\bp}_{\ell}, \bp_m})\}$, where $\alpha_{\hat{\bp}_{\ell}, \bp_m}$ is the angle between $\hat{\bp}_{\ell}$ and $\bp_m$. It varies as $0\leq \alpha_{\hat{\bp}_{\ell}, \bp_m}\leq \pi$. As the angle increases, the cosine decreases from 1 to $-1$, thereby increasing $\|\hat{\bp}_{\ell}-\bp_m\|_2^2$ and decreasing $\exp\left(-\frac{\|\hat{\bp}_{\ell}-\bp_m\|_2^2}{2h^2}\right)$.

Here, we model the logarithm of the normalized diffusion signal with respect to the $b=0$ images. The voxel-wise normalization using $b=0$ implicitly helps to remove intensity variations due to T2-weighting and inhomogeneity in response function \citep{tournier2019mrtrix3}. On the other hand, the log-scaling allows us to model the unknown function $f_{v}(\cdot,\cdot)$ to be unrestricted, thereby simplifying the model implementation.
Since there are only a handful of unique b-values, we use a polynomial basis expansion with respect to $b_m$'s.
However, the coefficients, $\theta_{v,k}(\bp_m)$'s, are assumed to vary with b-vectors $\bp_m$ as well as $k$, the order of the polynomial.
We model the $\theta_{v,k}(\bp_m)$'s as linear combinations of Gaussian kernels.
In the functional data analysis (FDA) literature, such an approach belongs to the class of radial basis function (RBF) based modeling of functional parameters.
RBF bases form an RBF network, a type of feed-forward neural network that is easier to work with for modeling multivariate functions than spline or other traditional univariate bases.
Furthermore, they enjoy universal approximation properties \cite{park1991universal, liao2003relaxed}, which ensure flexibility.
For simplicity, we only focus on Gaussian RBF bases in our analysis.
Furthermore, we use tapered Gaussian kernels, which are defined as $\exp(-d^2/2h^2)I\{d<3h\}$.
Similar to the mean and variance parameters of the Gaussian distribution,
Gaussian kernels are fully specified by centers and bandwidth parameters.
In our model, the centers are $\{\hat{\bp}_1,\ldots,\hat{\bp}_{2N}\}$ and a fixed bandwidth parameter $h$.
We keep the basis kernels fixed for all the polynomial orders, however, vary the coefficients $\beta_{v,k,\ell}$'s.

Due to the antipodal symmetry of dMRI signal, we must have $f_{v}(b_m,\bp_m)=f_{v}(b_m,-\bp_m)$. 
In our model, this is ensured by setting $\hat{\bp}_i=-\hat{\bp}_{i+N}$ and $\beta_{v,k,i}=\beta_{v,k,i+N}$.
It is easy to verify that under this specification, antipodal symmetry holds.

\subsection{Connection to existing dMRI models}
Although our proposed model in \eqref{eq1} shares commonalities with popular models used in dMRI data analysis such as Diffusion Tensor Imaging (DTI) or Diffusion Kurtosis Imaging (DKI) and more generally the Linearly Estimated Moments provide Orientations of Neurites And their Diffusivities Exactly (LEMONADE) model \cite{kiselev2010cumulant,novikov2018rotationally}.
The estimates from these approaches are used for summarizing fiber properties.
However, the estimated coefficient functions $\theta_{v,k}(\bp)$'s from our model do not have any direct relation with the estimates from DTI or DKI.
The motivation of our paper is primarily to get a denoised and smooth dMRI signal which may then be passed through DTI, DKI, LEMONADE, or NODDI to estimate the fiber properties.
Our estimation pipeline is data-driven.
For example, we may not even pre-specify $K$, but tune it using cross-validation-type approaches, as our aim is to obtain the optimal functional approximation of the observed dMRI signal.
Thus, the estimated coefficients from our model should not be used for summarizing fiber properties.




\subsection{Computation}
Voxel-wise fully nonparametric estimation of our proposed model is computationally prohibitive.
We thus propose an efficient computation method that works remarkably well in all of our experiments. 
This requires us to pre-specify the number of bases $N$ and order $K$. For convenience, we prescribe some default values for these parameters based on our experiment results.
Since, $\hat{\bp}_i=-\hat{\bp}_{i+N}$, we only need $N$-many distinct centers. We set them as evenly distributed points on $\mathbb{S}^2$ from the spherical Fibonacci lattice point set. 
Rest of the centers are obtained using the relation $\hat{\bp}_i=-\hat{\bp}_{i+N}$. 
To ensure uniqueness, we require $-1<\hat{\bp}_i^T\hat{\bp}_j<1$ for $i\neq j$.

For a given $N$ and $K$, we can re-write $f_v(b_m,\bp_m)$ as $\mathbf{x}_m^T\boldsymbol{\beta}_{v}$, where we set $\mathbf{x}_m=[b_m G(1,m)+b_m G(1+N,m),\ldots,b_m G(N,m)+b_m G(2N,m),\ldots ,b_m^K G(1,m)+b_m^K G(1+N,m),\ldots,b_m^K G(N,m)+b_m^K G(2N,m)],$ where $G(\ell,m)=\exp\left(-\frac{\|\hat{\bp}_{\ell}-\bp_m\|_2^2}{2h^2}\right)I\{\|\hat{\bp}_{\ell}-\bp_m\|_2<3h\}$ since $\beta_{v,k,i}=\beta_{v,k,i+N}$.

Similarly, $\boldsymbol{\beta}_{v} \in \mathbb{R}^{NK}$ be $\{\beta_{v,1,1},\ldots,\beta_{v,1,N},\beta_{v,2,1},\ldots,\beta_{v,2,N},\ldots,\beta_{v,K,1},\ldots,\beta_{v,K,N}\}$.
Varying above specification over $m$, we can build a $M\times NK$ design matrix $\mathbf{X}$ with $m$-$th$ row being set to $\mathbf{x}_m^T$. Thus, our proposed model simplifies to $\log(\mathbf{S}_{v})=\mathbf{X}\boldsymbol{\beta}+\mathbf{e}$, where $\mathbf{S}_{v}=\{S_{v}(b_1,\bp_1),\ldots,S_{v}(b_M,\bp_m)\}$. Therefore, it reduces to a linear regression model if $N$ and $K$ are not too large. In our experiments, setting $N=10$ and $K=4$ turns out to be adequate to get good numerical performances. It is however possible to set voxel-wise different orders based on some model selection criteria, such as Akaike information criterion (AIC) or Bayesian information criterion (BIC). 
The bandwidth parameter $h$ is set as $\frac{\sum_{i,j:a_{i,j}\neq 0} a_{i,j}}{\sum_{i,j} 1\{a_{i,j}\neq 0\}}$, where $a_{i,j}=\sqrt{2}\|\hat{\bp}_i-\hat{\bp}_j\|_2$.

The above simplification leads us to estimate $\boldsymbol{\beta}_v$ using ordinary least square and obtain $\boldsymbol{\hat{\beta}}_v=(\mathbf{X}^T\mathbf{X})^{-1}\mathbf{X}\mathbf{y}_v$, where $\mathbf{y}_v=\log(\mathbf{S}_{v})$. Since $(\mathbf{X}^T\mathbf{X})^{-1}\mathbf{X}^T$ is independent of $v$ and thus is not needed to be calculated for each voxel location separately.
We need to compute the above matrix only once and then multiply it with the transformed dMRI signal to estimate the coefficients.
However, to avoid any numerical instability, we compute $(\mathbf{X}^T\mathbf{X}+d\mathbf{I}_{NK})^{-1}\mathbf{X}^T$ for some small scalar $d>0$ and $\mathbf{I}_{NK}$ is an identity matrix of dimension $NK\times NK$.
We perform all analyses and coding in R \citep{Rprogramming} using {\tt Rcpp} \citep{Eddelbuettel2011Rcpp:Integration} and {\tt RcppArmadillo} \citep{Eddelbuettel2014RcppArmadillo:Algebra}. 

\subsection{Applications}
\label{sec:application}

Our proposed model [\ref{eq1}] primarily helps to de-noise the diffusion data and obtain a smooth approximation of the diffusion signal as a function of gradient direction and gradient strength.
There are several usages of such representation.
In this paper, we show two applications, namely 1) signal prediction and 2) data harmonization.

\subsubsection{Diffusion Signal Prediction}
The dMRI preprocessing steps (e.g., the steps in the FSL {\tt eddy} toolbox) heavily rely on a good diffusion signal prediction algorithm  for 1) estimating motion for any frame (dMRI data obtained at given $(b_m,\bp_m)$); and 2) for removing and replacing outlier frames.
We compare the out-of-sample prediction performance of our method with the GP model in FSL {\tt eddy} and the RBF model in \cite{ning2015estimating}. In particular, we compute the L2 norm of the difference between a measured and the predicted signal in the log-scale in test data. 


\subsubsection{Diffusion Microstructural Feature Harmonization}
\label{sec:harmonization}
In dMRI, acquisition protocol variations can affect the resulting microstructural features in the brain, including commonly used measures such as  diffusion tensor imaging (DTI) and NODDI metrics \cite{zhang2012noddi}. These features are derived non-linearly from dMRI data and can be influenced by the specific b-value and b-vector design used in each study. For instance, the Human Connectome Project Young Adults (HCP-YA) study utilized three b-values or shells (b = 1000, 2000, and 3000) with 90 b-vectors on each shell \cite{glasser2016human}, while the Adolescent Brain Cognitive Development (ABCD) Study employed four shells, with 6 b-vectors on b = 500, 16 on b=1000, 16 on b=2000, and 60 on b=3000 \cite{casey2018adolescent}. Therefore, when comparing microstructure features across different studies, it is crucial to harmonize these measures as a preprocessing step. However, existing harmonization techniques such as ComBat are based on linear models and may not be suitable for handling non-linear batch effects resulting from differences in acquisition protocols. Furthermore, ComBat is only suitable to apply on post-processed summarized features such as FA, but not the diffusion signal.

\begin{figure}
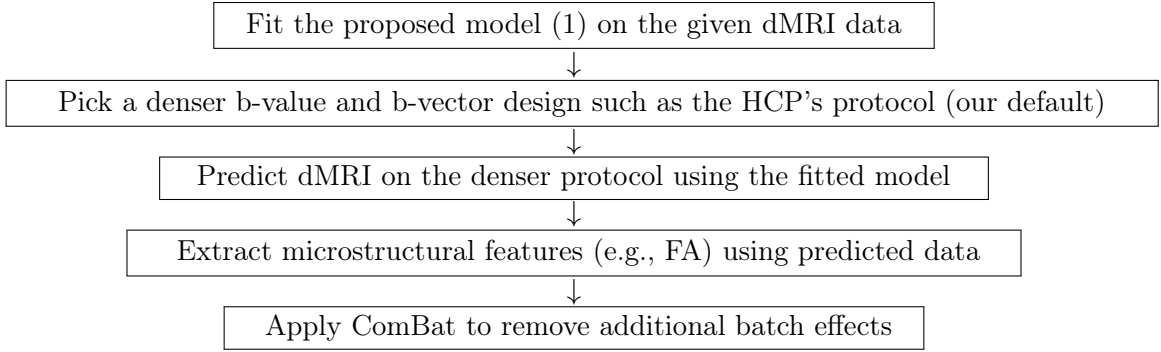

\begin{center}
\framebox[1.1\width]{Fit the proposed model (\ref{eq1}) on the given dMRI data} \par $\downarrow$ \par
\framebox[1.1\width]{Pick a denser b-value and b-vector design such as the HCP's protocol (our default)}
\par $\downarrow$ \par
\framebox[1.1\width]{Predict dMRI on the denser protocol using the fitted model}\par $\downarrow$ \par
\framebox[1.1\width]{ Extract microstructural features (e.g., FA) using predicted data}\par $\downarrow$ \par
\framebox[1.1\width]{ Apply ComBat to remove additional batch effects}
\end{center}
\caption{Diffusion MRI microstructural feature harmonization through signal prediction.}
\label{fig:dataharm}
\end{figure}

In this section, we design a data harmonization technique using our diffusion prediction method to account for non-linear effects caused by acquisition variations.
Specifically, the proposed harmonization pipeline consists of five main steps: 1) Fit the proposed model (\ref{eq1}), on the given dMRI data. 2)
Choose a denser b-value and b-vector design, such as the Human Connectome Project's (HCP) protocol, which is used as the default in this study. 3)
Predict dMRI on the denser protocol using the fitted model. 4)
Extract microstructural features, such as Fractional Anisotropy (FA), using the predicted data. And 5)
apply the ComBat technique to remove any additional batch effects that may be present. For clarity, the detailed steps are described in Figure \ref{fig:dataharm}.

{To enhance the quality and consistency of diffusion signals, steps 1-4 are designed to standardize the b-values and b-vectors. Despite these efforts, variations in scanner hardware and other external factors can introduce additional batch effects into the standardized diffusion data and the derived microstructural features. To counter these effects and ensure higher data quality across multiple scanners or sites, we employ the ComBat method on the extracted microstructural features. This approach is effective in further reducing batch effects. Note that our method works best when the maximum b-value in step 2 is not more than the maximum b-value of the original data.}


\subsubsection{Data Sets}
We use two data sets to demonstrate the effectiveness of our method. The first data set is derived from the HCP. The second originates from the Multi-shell Diffusion MRI Harmonisation Challenge (MUSHAC), 2018. The details of the two datasets are as follows.

{\bf Human Connectome Project (HCP) Data:}
The HCP is a large data set consisting of dMRI data gathered from approximately 1,200 healthy adults. Customized scanners were utilized to generate high-quality and consistent data for measuring brain connectivity. The data can be easily accessed at https://db.humanconnectome.org/. 
A full dMRI session in HCP includes 6 runs, each lasting approximately 10 minutes. These runs include three distinct gradient tables, with each table acquired once using right-to-left and left-to-right phase encoding polarities, respectively. Each gradient table includes approximately 90 diffusion-weighted directions along with six $b=0$ acquisitions interspersed throughout each run. Diffusion weighting consisted of 3 shells of $b=1000, 2000$, and $3000$. The scan was conducted using a spin-echo EPI sequence on a 3 Tesla customized Connectome Scanner and the dMRI data have an isotropic resolution of 1.25 $mm$. For more details about HCP data acquisition, please refer to \cite{VanEssen20122222}.

{\bf Multi-shell Diffusion MRI Harmonisation Challenge (MUSHAC) Data:}
The MUSHAC data set offers cross-scanner and cross-protocol multi-shell  dMRI data \cite{tax2019cross,ning2020cross}. We downloaded data for 10 subjects, with each subject undergoing four scans from two scanners: a) the 3T Siemens Prisma (80 $mT/m$) and b) the 3T Siemens Connectom (300 $mT/m$), using two distinct protocols. These protocols comprise: 1) a conventional (ST) protocol with acquisition parameters conforming to a standard clinical protocol; and 2) an advanced (SA) protocol that capitalizes on the enhanced hardware and software specifications to augment the number of acquisitions and spatial resolution per unit time. The mean interval between acquisitions on scanners a) and b) was one month, with no software upgrades occurring during the study.

The ST data from both scanners have an isotropic resolution of 2.4 $mm$ and 30 directions at $b = 1200$ and $3000$. The Prisma-SA data has a superior isotropic resolution of 1.5 $mm$ and 60 directions at the same b-values, while the Connectom-SA data have the highest resolution of 1.2 $mm$ and 60 directions. All protocols incorporated reversed-phase encoding, and multiple b0 images.


\section{Results}
\label{sec:exp}
We ran three experiments to highlight the utilities of the proposed method.
The first experiment was on predicting the diffusion-weighted signal and its comparison with the state-of-the-art methods.
After that, we showed the harmonization application of the proposed method in two different settings. The first one evaluated cross-protocol harmonization and the other one studied cross-protocol and cross-scanner harmonization

\subsection{Experiment 1: Diffusion-Weighted Signal Prediction}

{ Accurate prediction of the dMRI data on q-space is an important step for several applications such as data standardization, denoising against outliers, signal prediction for phantom data generation, etc. The prediction problem is particularly hard as the available data is usually observed on a dense grid of b-vectors, but an extremely sparse grid of b-values. This experiment is thus structured to compare the predictive performance of the proposed method against some of the existing alternatives.}  We randomly chose one subject's dMRI data from the HCP. The data contains 270 diffusion-weighted images, which were partitioned into training and testing sets at a ratio of 75:25. Accordingly, approximately 203 ($270 \times 0.75 \approx 203$) images were allocated to the training set and 67 were allocated to the testing set. These training images were further subsampled to emulate sparser protocols than the HCP one. {Given that each subject encompasses millions of voxels, and each voxel comprises 270 diffusion measures, the results presented here, derived from a subject selected at random, are sufficiently representative of our method's performance across other subjects.}

We conducted experiments with six distinct designs or protocols, as delineated in Table~\ref{tab: protocol}. Following each protocol's design, 50 samples were drawn from the training diffusion-weighted images, resulting in 50 replicas of the training set, each containing 105 diffusion-weighted images. Our model was subsequently fitted to the training data together with 18 b0 images.
Finally, the signal was predicted at the (b-value, b-vector) pairs present within the test set.

\begin{table}[htbp]
    \centering
    \caption{A single-subject's HCP data was resampled based on the following protocols with different numbers of gradients on each shell. Under each protocol, 50 random samples were considered.}
    \begin{tabular}{crrrc}
    \toprule
    \multicolumn{5}{c}{Experiment 1 protocol settings}\\
    \midrule
    Protocol & $b=$1000 & $b=$2000 & $b=$3000 & Replications \\
    \midrule
    1 & 60 & 30 & 15 & 50 \\
    2 & 60 & 15 & 30 & 50 \\
    3 & 30 & 60 & 15 & 50 \\
    4 & 30 & 15 & 60 & 50 \\
    5 & 15 & 30 & 60 & 50 \\
    6 & 15 & 60 & 30 & 50 \\
    \bottomrule
    \end{tabular}
    \label{tab: protocol}
\end{table}

We compared our model with two popular methods: 1) the GP method in FSL \cite{andersson2015non} and 2) the RBF-based method in \cite{ning2015estimating}. The GP method was implemented in the {\tt Eddy} tool in FSL, and its prediction applies to each slice (along the third dimension in a brain volume).  
Hence, we picked the 60-$th$ brain slice, which has a good mixture of different types of brain tissue. To evaluate the GP model in FSL, we relied on its outlier detection mechanism which simultaneously detects outlier slices and replaces them with GP-based predictions. In each run, we built dMRI data with 105 training images, 67 testing images and 18 b0 images. For the 67 test images, we replaced its value with a constant (i.e., $M_v-10s_v$, where $M_v$ and $s_v$ are the mean and standard deviation of the signal from the 105 training images at voxel $v$) to ensure that those were detected as outliers. We also ensured that none of training images were detected as outliers. The samples used as training data by FSL remain the same in {\tt Eddy} corrected data, which helped us verify whether training images in FSL are actually what we were aiming for. On the other hand, checking accurate detection of the test images is straightforward. 
We reported results based on a subset of voxels where the test images were detected correctly, but all intended training images were still identified correctly. However, this subset constitutes more than 90\% of all the voxels for all the protocols. 
To assess the RBF-based method in \cite{ning2015estimating}, we first scaled the data by dividing it with the average b0 signal value. We then fitted the model using the training data. Moreover, the RBF method \cite{ning2015estimating} is very time-demanding, and thus, we reduced our computing to a subset of $\sim$ 400 voxels on slice-60 while comparing all three methods. However, the results evaluated for the entire 60-$th$ slice are also provided, but only for FSL and our proposed method in Table~\ref{tab: tab2full}. { Since all of these methods are voxel-wise, the test sets are reasonably large.}

{ For our model comparison, the mean squared error (MSE) between predicted and measured signals at the log scale is considered, as it provides a model-agnostic assessment of the fit.} The MSEs are reported in Tables \ref{tab: tab2} and \ref{tab: tab2full}. We can see that our method performs the best among the three compared methods, except for Protocol 2 in Table \ref{tab: tab2}. In Protocol 2, \cite{ning2015estimating} performs the best. In Table~\ref{tab: tab2full} however, prediction of our method consistently beats the FSL-based prediction. In addition, our method runs much faster than {\tt Eddy} even when using {\tt Eddy\_cpu} or {\tt Eddy\_openmp} and the RBF method in \cite{ning2015estimating}. 
The basic architecture of {\tt Eddy} with {\tt --repol} flag is to identify the outliers and then replace them with the predictions made by the GP model.
Specifically, it runs an iterative procedure combining the movement corrections with the outlier replacements.
Thus, the computation time is not directly comparable with our method.
However, for the latter method, we could directly predict the signal after estimating the model parameters using the codes shared in \cite{ning2015estimating}. Although this method shares some commonalities with our proposed approach, it is computationally much more demanding than ours.
For the computation with 400 voxels, it took around 10 mins for \cite{ning2015estimating}. In contrast, our approach requires only one matrix-vector multiplication per voxel, and it takes only a few seconds to compute signals for the same number of voxels. And it takes about a minute to fit our model for all voxels in the whole brain with a dense HCP dMRI protocol. These results demonstrate that our proposed approach is a fast and effective way to denoise dMRI data, which can facilitate the analysis of large-scale studies.

\begin{table}[ht]
\centering
\footnotesize
\caption{Prediction MSEs under different sparse protocols (the signal was log-transformed) for the subset of 400 voxels. The best performance for each protocol is bolded.}
\begin{tabular}{rrrrrrr}
  \hline
 & Protocol-1 & Protocol-2 & Protocol-3 & Protocol-4 & Protocol-5 & Protocol-6 \\ 
  \hline 
FSL &  0.23   &    0.45& 0.37 & 0.48 &0.46   &    0.44\\ 
Ning et al. (2015) \cite{ning2015estimating} &0.30 &    {\bf 0.30}  &  0.39  &  0.44  &  0.45 &   0.40 \\
Poly-RBF$(4,10)$  & {\bf 0.21} &0.36 &{\bf 0.35} &{\bf 0.37} &{\bf 0.36} &{\bf 0.35}\\  
   \hline
\end{tabular}
\label{tab: tab2}
\end{table}

\begin{table}[ht]
\centering
\footnotesize
\caption{Similar to Table \ref{tab: tab2} but with MSEs computed for voxels in the entire 60-th slice. The method described in \cite{ning2015estimating} is excluded due to its high computational intensity.}
\begin{tabular}{rrrrrrr}
  \hline
 & Protocol-1 & Protocol-2 & Protocol-3 & Protocol-4 & Protocol-5 & Protocol-6 \\ 
  \hline 
FSL &  0.25 & 0.48 & \bf{0.35} & 0.49 & 0.43 & 0.41 \\ 
 Poly-RBF$(4,10)$ & \bf{0.17} & \bf{0.37} & \bf{0.35} & \bf{0.34} & \bf{0.31} & \bf{0.27} \\ 
   \hline
\end{tabular}
\label{tab: tab2full}
\end{table}


\subsection{Experiment 2: Cross-Protocol Harmonization}

In this section, we evaluated our model on cross-protocol dMRI data harmonization, where the data were acquired from the same scanner but using different settings of b-values and b-vectors. We re-used the re-sampled data from the previous section, containing 6 different acquisition protocol designs and 50 samples for each design, for validation. The primary focus of this study was on the harmonization of NODDI \citep{zhang2012noddi} metrics. 
The NODDI metrics were preferred because their estimation utilizes multi-shell dMRI data. 
Specifically, we focused on the orientation dispersion index (ODI).

We mainly evaluated whether the proposed harmonization strategy in Section \ref{sec:harmonization} can harmonize the derived NODDI metrics. Since our data are from the same scanner and subject, we did not include the ComBat step (step 5) for this experiment. For each set of simulated dMRI data, we got two sets of NODDI metrics: 1) the NODDI metrics by directly applying the NODDI model to the original resampled dMRI data, and 2) the NODDI metrics by applying the NODDI model to the pipeline-harmonized (Section \ref{sec:harmonization}) resampled dMRI data.

In 1), although all our data were subsampled from one HCP subject, the estimated NODDI metrics are different due to different acquisition designs. We first ran two-sample paired t-test at each voxel to compare the difference for each pair of protocols before (50 samples) and after dMRI harmonization (50 samples). In Figure \ref{fig:harmo1}, we use the comparison between Protocol 1 v.s. Protocol 3 as an illustrative example in the main paper (other comparisons can be found in the Supplement). Larger values of $-log(\text{p-values})$ imply bigger deviations between protocols. We used different thresholds to sort out the voxels which present larger deviations between protocols. In the upper panel of Figure \ref{fig:harmo1}, we show the number of voxels whose $-log(\text{p-values})$ that are larger than a sequence of thresholds. Naturally, for larger cutoffs on $-log(\text{p-values})$, fewer number of voxels are identified as different. However, we can see that the survived number of voxels are always smaller in the “harmonized” case than in the “original” case, indicating the harmonized case has smaller difference on the computed ODI values between Protocol 1 and Protocol 3. In the second row, we visualize the voxel-wise $-log(\text{p-values})$ for voxels whose p-values are smaller than 0.05. Here, we can clearly see that harmonized case has much smaller differences (i.e., smaller $-log(\text{p-values})$) than the original case.

Comparisons for other pairs of subsampled data can be found in the Supplementary Materials (Figure S.1-S.14). In all the comparisons, we can see that our harmonization pipeline substantially reduced the difference between different protocols.



\begin{figure}
    \centering
    \includegraphics[width = 1\textwidth]{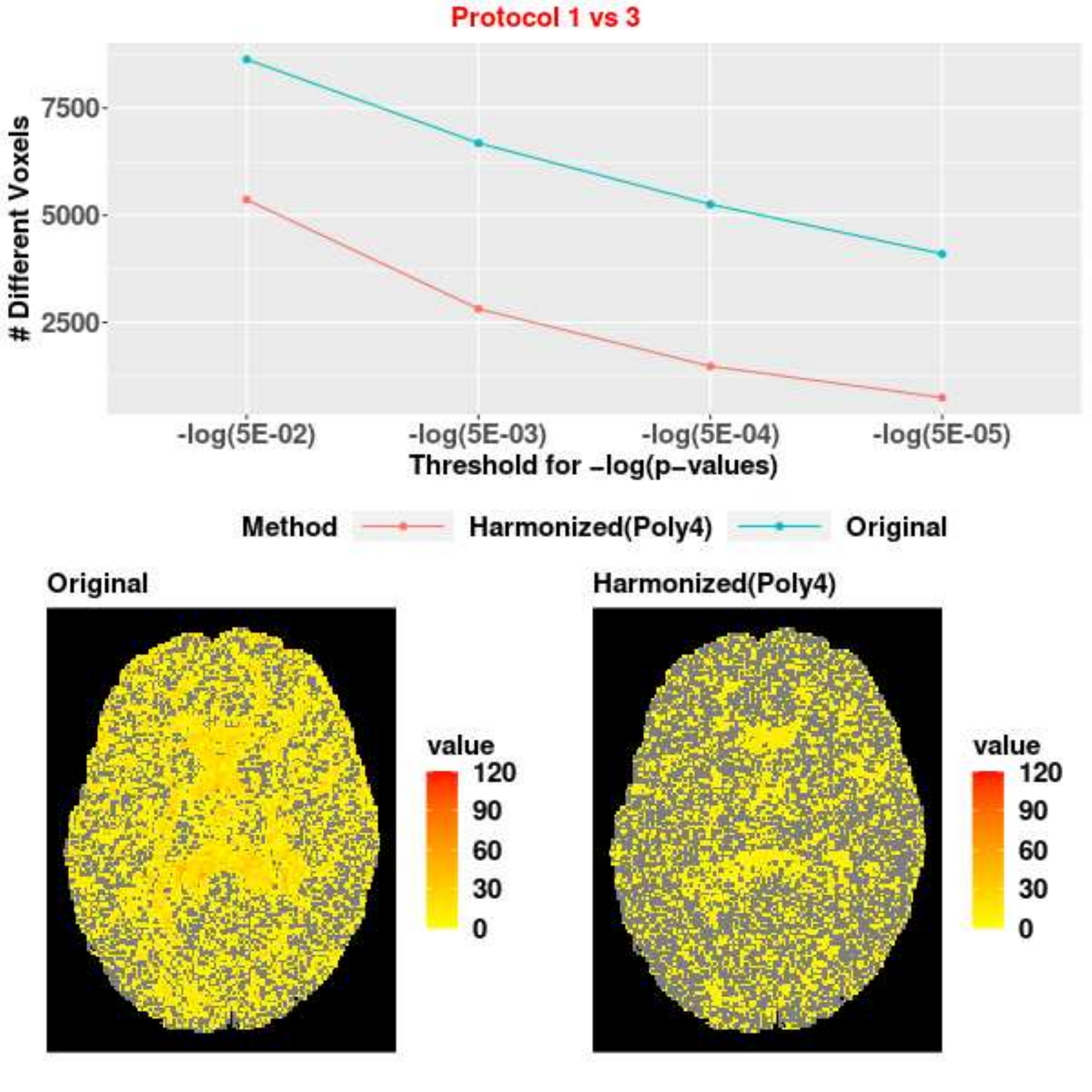}
    \caption{Evaluation of the difference between the ODI values produced by Protocols 1 and 3 (Table \ref{tab: protocol}). The first row shows the number of voxels whose $-log(\text{p-values})$ (obtained from two-sample paired t-tests) are smaller than a sequence of thresholds. The second row presents voxel-wise $-log(\text{p-values})$ of the t-test for only voxels with $-log(\text{p-values})$>-log(0.05) (i.e., the voxels of differences). The harmonized data was obtained by applying our proposed model with $K = 4$.
    }
    \label{fig:harmo1}
\end{figure}

To provide another perspective in evaluating the performance of harmonization, we compared the original and harmonized ODI values with a \textit{gold standard} ODI, computed from the full data, i.e., all 270 (b-value, b-vector) pairs in the HCP data. 
We present the mean absolute ODI difference maps between the gold standard and original and harmonized cases under Protocol 1 in the second row of Figure \ref{fig:absdiff1}. In the first row, we show the values of absolute difference for different quantiles of a certain method’s absolute difference; the ``harmonized'' case shows an overwhelmingly better performance at larger quantiles. The results of absolute differences of other protocols are given in the supplementary materials. The results of absolute values for other protocols are in the Supplementary Materials (Figure S.15-S.19). All the results imply a consensus conclusion – the proposed harmonization approach reduces the extreme differences between the ODI values derived from the subsampled data and the gold standard ODIs derived from the full HCP data.


\begin{figure}
    \centering
    \includegraphics[width = 1\textwidth]{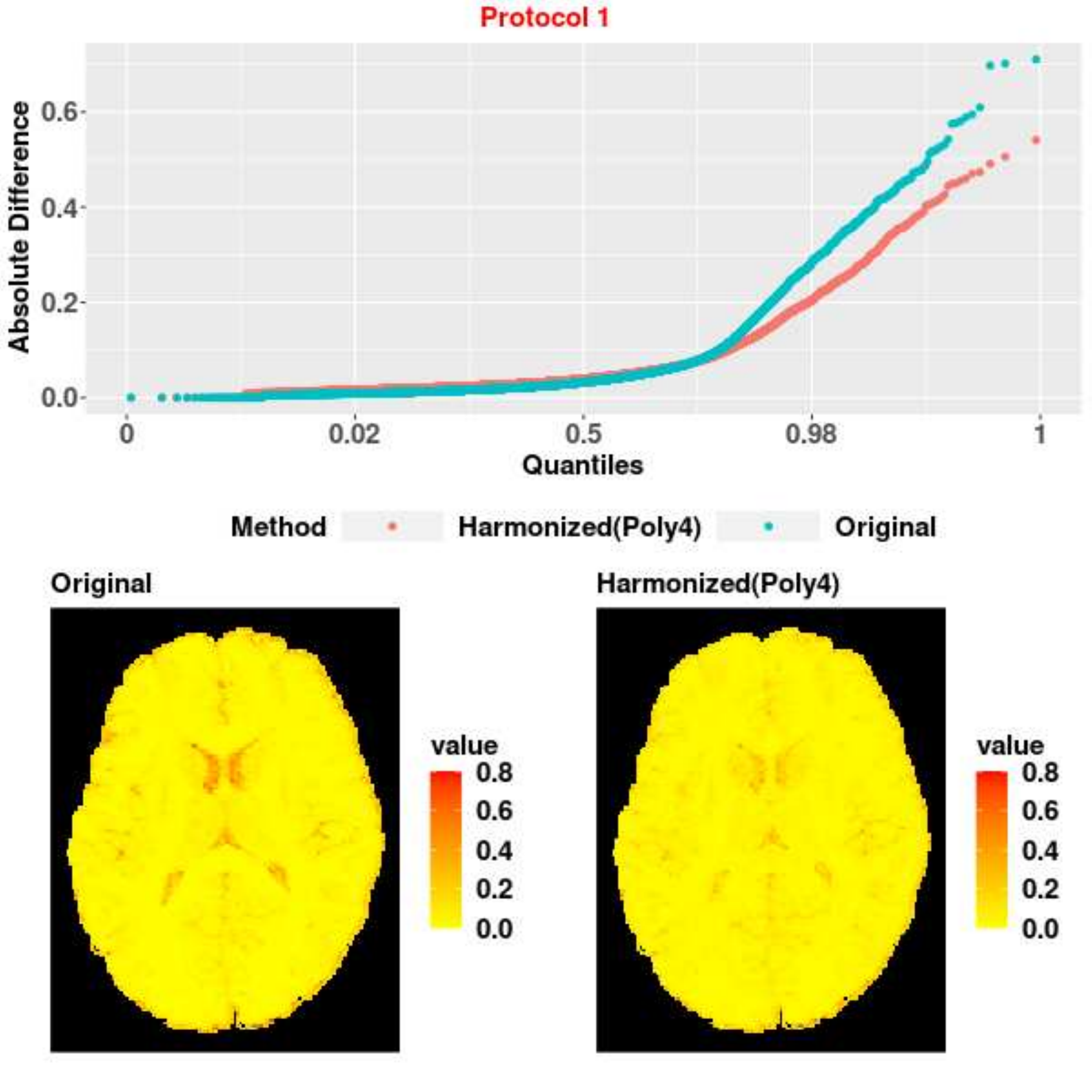}
    \caption{
    The first row compares the quantiles of the mean absolute difference of ODI values between the gold standard one and those obtained from the original and harmonized re-sampled dMRI data. The second row shows maps of the absolute difference between the gold stand ODI values and the original signals and harmonized ones. The harmonized data is obtained by applying our proposed method with $K = 4$.
    }
    \label{fig:absdiff1}
\end{figure}

\subsection{Experiment 3: Cross-Scanner and Cross-Protocol Harmonization}
Large-scale neuroimaging studies are often conducted via multiple centers and thus the data is collected using multiple scanners.
This leads to potential variations in the downstream analysis due to scanner-effect \citep{wittens2021inter}.
The situation becomes more concerning when different centers use different diffusion protocols. In this section, we  examined the utility of our proposed harmonization technique under a cross-scanner, cross-protocol regime.
The MUSHAC data was used as it contains cross-protocol and cross-scanner data. Here we focused on the ODI metric. And all ODI images were registered to the MNI space for downstream statistical analysis.

We first applied the first four steps of the proposed harmonization protocol described in Figure \ref{fig:dataharm}.
In step 1, we fitted the proposed model (\ref{eq1}) using $K=4$, and in step 4 we passed the predicted (Harmonized) diffusion signals to the NODDI model to obtain the ODI metric. {In step 5, we applied ComBat for additional harmonization.} In addition, we estimated the ODI metric from the original data as well (denoted as `Original'). 

Using the MUSHAC data, we could compute 4 ODI images for each subject, as the data were acquired using two acquisition protocols under two scanners.
We evaluated the reproducibility of ODI features using intra-class correlation (ICC) (see \cite{liljequist2019intraclass}), which is defined as $(1-\hat\sigma^{2}_{\text{within--subject}}/\hat\sigma^{2}_{\text{total}})$, where $\hat\sigma^{2}_{\text{within--subject}}$ stands for the variability across different protocols and scanners within a given subject and $\hat\sigma^{2}_{\text{total}}$ is the overall variability.
ICC ranges between (0,1), and a higher value corresponds to higher reproducibility of the cross-protocol and cross-scanner data. We computed the voxel-wise ICC for the ODI metric obtained from the original data, and compared it with the ICC obtained from the Poly-RBF harmonized dMRI with and without the ComBat step. 
{ Specifically, we are comparing four sets of ICC estimates for ODI maps extracted from 1) the original data directly, 2) the original data and subsequently combat-adjusted, 3) Poly-RBF harmonized data, and 4) Poly-RBF harmonized data and subsequently combat-adjusted. We stratify these four cases by directly comparing case 1 with case 3 and then case 2 with case 4 to see if Poly-RBF can improve reproducibility with and without Combat. In this context, comparisons between cases 1 and 3 are referred to as `Without ComBat', and those between cases 2 and 4 are termed `With ComBat'.}

\begin{figure}[ht!]
    \centering
    \includegraphics[width=1\textwidth]{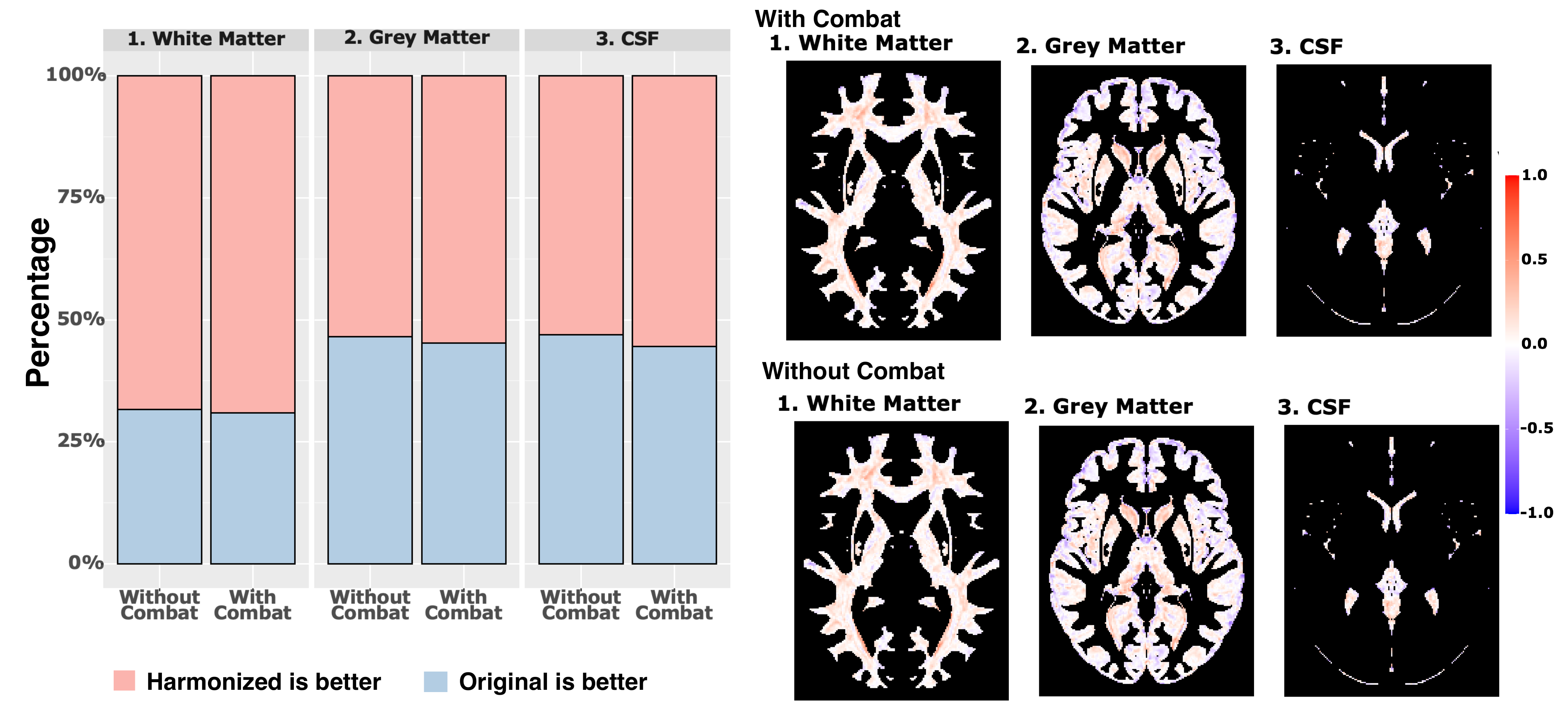}
\caption{ODI reproducibility study using the proposed Poly-RBF harmonization method. The left panel displays percentage-stacked bar plots illustrating the voxel proportions exhibiting improved ICC for Poly-RBF harmonized data (in orange) compared to the original data, alongside proportions showing deteriorated ICC (in light blue). The right panel depicts ICC difference maps (Harmonized $-$ Original) both with and without ComBat adjustments. Areas of improvement are highlighted in red, whereas areas of deterioration are shown in blue. The analyses are conducted separately for White Matter, Grey Matter, and CSF regions.}

    \label{fig:exp4}
\end{figure}

In Figure \ref{fig:exp4}, we present our findings individually for the White Matter, Grey Matter, and CSF regions to illustrate the varying patterns in the ICC results. For each type of region, we have two distinct sets of results: one with ComBat applied (cases 2 vs. 4) and one without (cases 1 vs. 3). In the left section of this figure, we also show that the ICC for harmonized data (with Poly-RBF applied) is superior, reaching approximately 70\% in the entire white matter regions. For grey and CSF regions, however, the percentages are nearly 55\%. The `With ComBat' results are also better than the `Without ComBat' cases, suggesting that both Poly-RBF harmonization and ComBat should be applied in harmonizing microstructural features. The right part of this figure displays the difference maps of the ICC (Harmonized $-$ Original) with and without ComBat adjustments separately. Thus, a positive number at a given location suggests that the ICC of harmonized data (with Poly-RBF applied) is better than the ICC of the original data at that location. And a negative number would imply the opposite. We plot ICC-difference results for the white matter, grey matter, and CSF regions of the 60-th brain slice. We observe that most of the white-matter regions are either white or reddish, implying better ICC of the harmonized data. This finding suggests that our Poly-RBF harmonization method is effective in significantly enhancing the reproducibility of the ODI measurement obtained from different scanners and protocols.


\section{Discussion}
\label{sec:discussion}
We present a novel and efficient algorithm for modeling dMRI signals in the q-space. Given noisy dMRI measures on an extremely sparse grid of b-values and a moderately dense grid of gradient directions, we can accurately estimate normalized diffusion signal $S_v(b,{\bf p})$ as a function of b-values and b-vectors. We demonstrate the superior predictive ability of our model over existing state-of-the-art methods, including the popular GP model implemented in the {\tt Eddy} tool, using subsampled HCP data. Additionally, we apply our method to harmonize dMRI data acquired from different scanners and protocols and show that our approach leads to improved reproducibility of diffusion metrics (i.e., the ODI) computed from the harmonized data over the original data. The implementation of our model is available in {\url{https://github.com/royarkaprava/Poly-RBF_code}}. 

However, there are opportunities to improve the proposed method. To increase efficiency, we pre-specify the hyperparameters in the model, such as the RBF basis functions and the polynomial degree $K$. While the pre-specified parameters may not be optimal for each voxel, our approach still achieves improved prediction compared to competitors. A future improvement would be to incorporate spatially varying hyper-parameters to optimize the model for each voxel. Furthermore, the current model may have limited predictive power for $(b,{\bf p})$ pairs outside of the training data's b-value range. In other words, when data is only collected for $b = 1000$ and $2000$, predictions for the signal at $b=4000$ may be inaccurate. In terms of applications, we primarily focus on diffusion signal harmonization. Our proposed method provides great computational efficiency and high prediction accuracy, making it suitable for quickly inferring motion and outlier frameworks in dMRI, which will be explored in the future.

\section*{Data Availability Statement
}
The open-source code is available at {\url{https://github.com/royarkaprava/Poly-RBF_code}}. For our analysis, publicly available datasets are used, as described in the main text.

\section*{Author Contributions}
A.R.: conceptualization, data curation, investigation, methodology, project administration, software, validation, writing—original draft, and writing—review \&
editing; Z.L.: conceptualization, data curation, methodology, validation, visualization, writing—original draft, and
writing—review \& editing; Z.Z.: conceptualization,data curation, methodology, supervision, visualization, writing—original draft, and writing—review \&
editing; 

\section*{Declaration of Competing Interests}
None of the authors has a conflict of interest to declare.

    \section*{Supplementary figures}
\begin{figure}
    \centering
    \includegraphics[width = 0.9\textwidth]{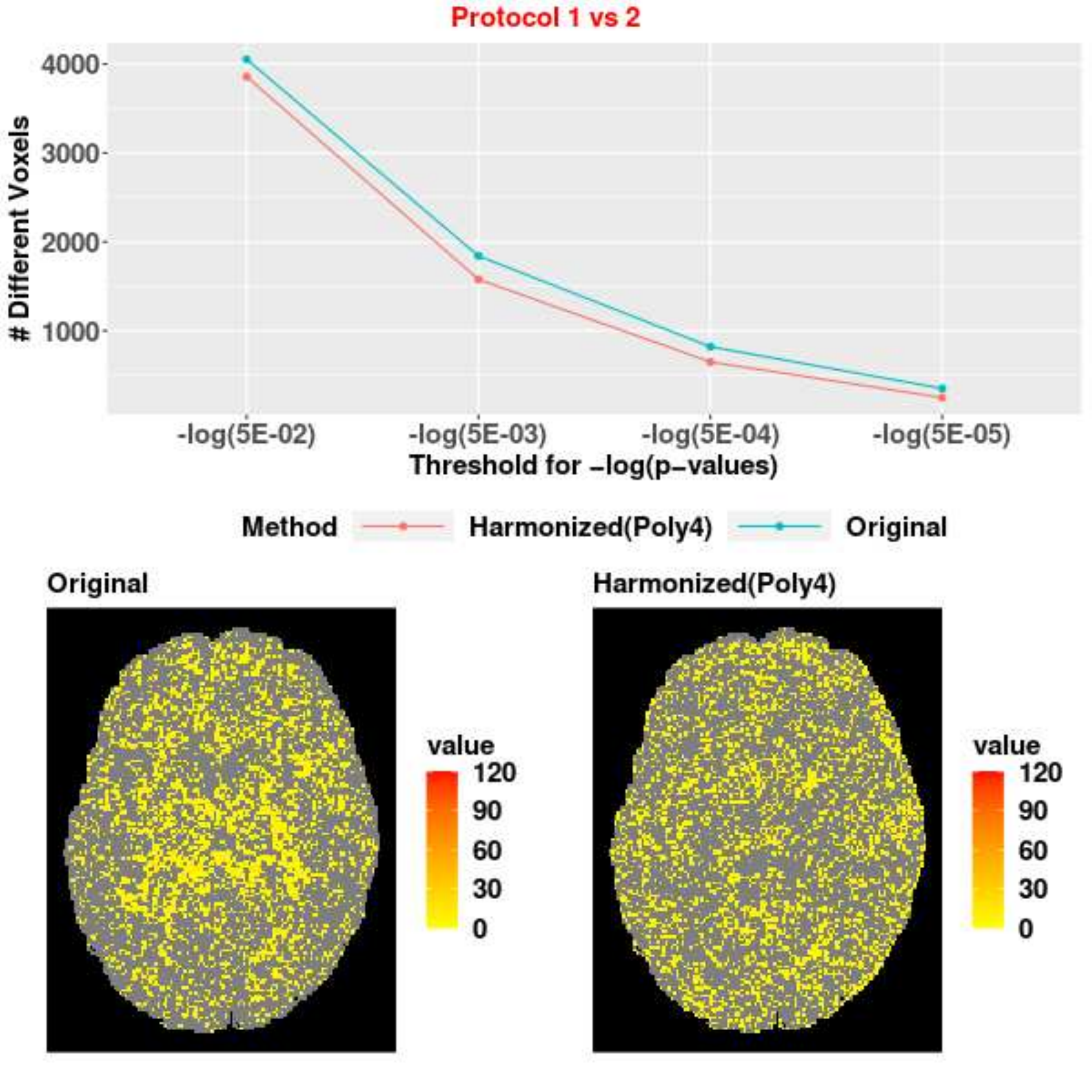}
    \caption{Evaluation of difference of NODDI metrics produced by protocols 1 and 2. The first figure on the first row shows the number of voxels with $-log(\text{p-value})$ of two-sample paired t-tests more than a given threshold.
    The other pair of images on the second row present voxel-wise $-log(\text{p-value})$ of two-sample paired t-tests. The harmonized data is obtained by applying our proposed model with $K=4$.
    }
    \label{fig:harmo1}
\end{figure}

\begin{figure}
    \centering
    \includegraphics[width = 0.9\textwidth]{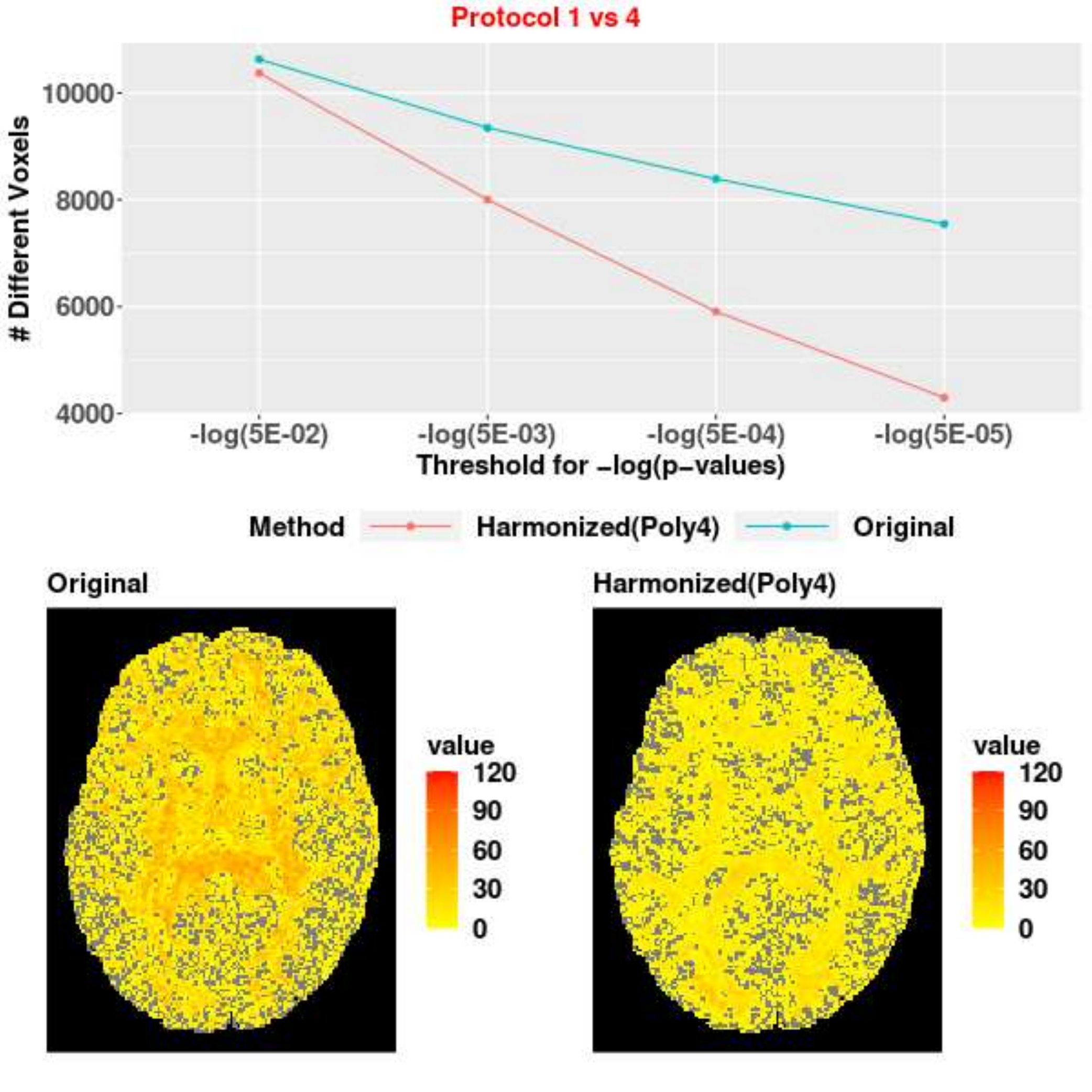}
    \caption{Evaluation of difference of NODDI metrics produced by protocols 1 and 4. The first figure on the first row shows the number of voxels with $-log(\text{p-value})$ of two-sample paired t-tests more than a given threshold.
    The other pair of images on the second row present voxel-wise $-log(\text{p-value})$ of two-sample paired t-tests. The harmonized data is obtained by applying our proposed model with $K=4$.
    }
    \label{fig:harmo1}
\end{figure}

\begin{figure}
    \centering
    \includegraphics[width = 0.9\textwidth]{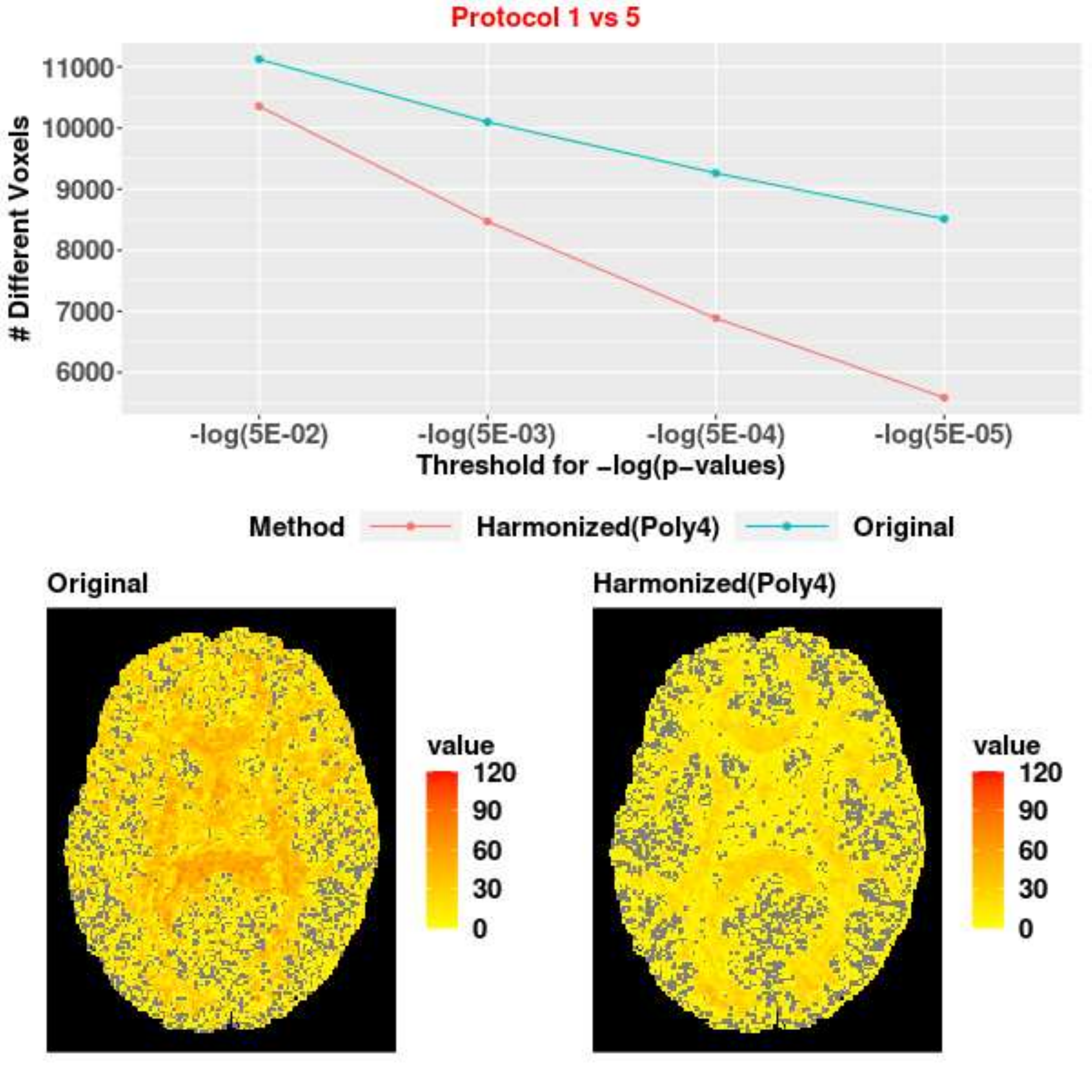}
    \caption{Evaluation of difference of NODDI metrics produced by protocols 1 and 5. The first figure on the first row shows the number of voxels with $-log(\text{p-value})$ of two-sample paired t-tests more than a given threshold.
    The other pair of images on the second row present voxel-wise $-log(\text{p-value})$ of two-sample paired t-tests. The harmonized data is obtained by applying our proposed model with $K=4$.
    }
    \label{fig:harmo1}
\end{figure}

\begin{figure}
    \centering
    \includegraphics[width = 0.9\textwidth]{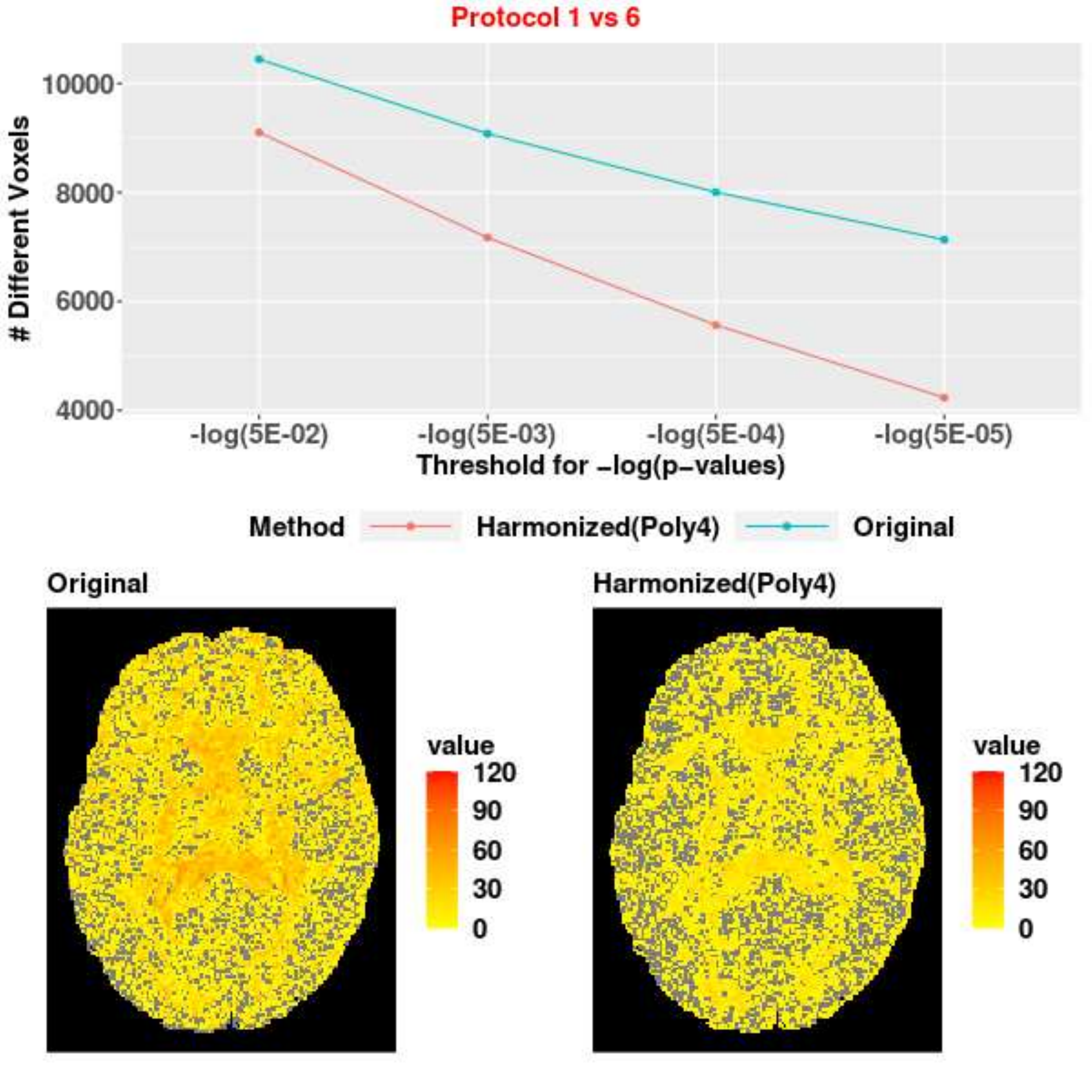}
    \caption{Evaluation of difference of NODDI metrics produced by protocols 1 and 6. The first figure on the first row shows the number of voxels with $-log(\text{p-value})$ of two-sample paired t-tests more than a given threshold.
    The other pair of images on the second row present voxel-wise $-log(\text{p-value})$ of two-sample paired t-tests. The harmonized data is obtained by applying our proposed model with $K=4$.
    }
    \label{fig:harmo1}
\end{figure}

\begin{figure}
    \centering
    \includegraphics[width = 0.9\textwidth]{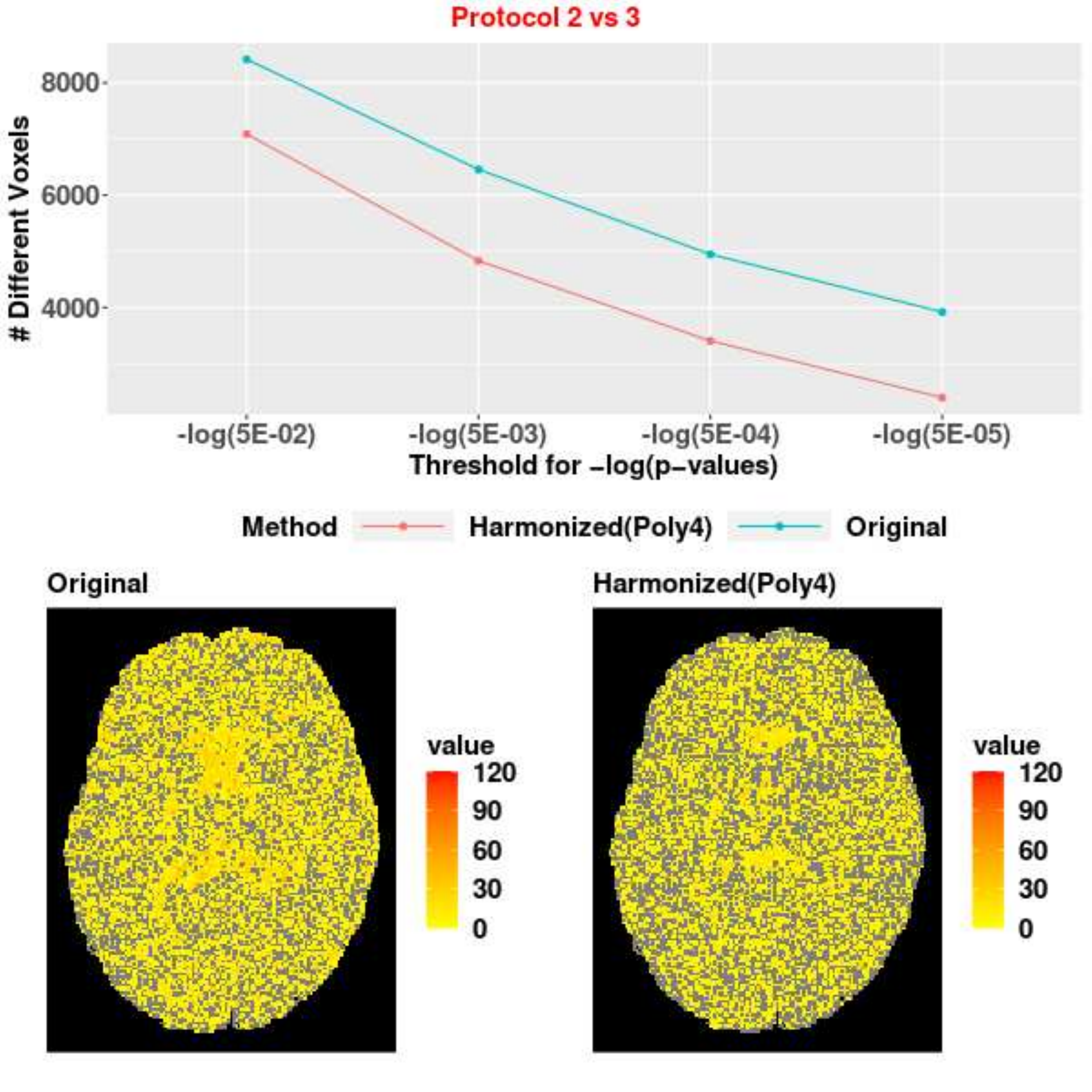}
    \caption{Evaluation of difference of NODDI metrics produced by protocols 2 and 3. The first figure on the first row shows the number of voxels with $-log(\text{p-value})$ of two-sample paired t-tests more than a given threshold.
    The other pair of images on the second row present voxel-wise $-log(\text{p-value})$ of two-sample paired t-tests. The harmonized data is obtained by applying our proposed model with $K=4$.
    }
    \label{fig:harmo1}
\end{figure}

\begin{figure}
    \centering
    \includegraphics[width = 0.9\textwidth]{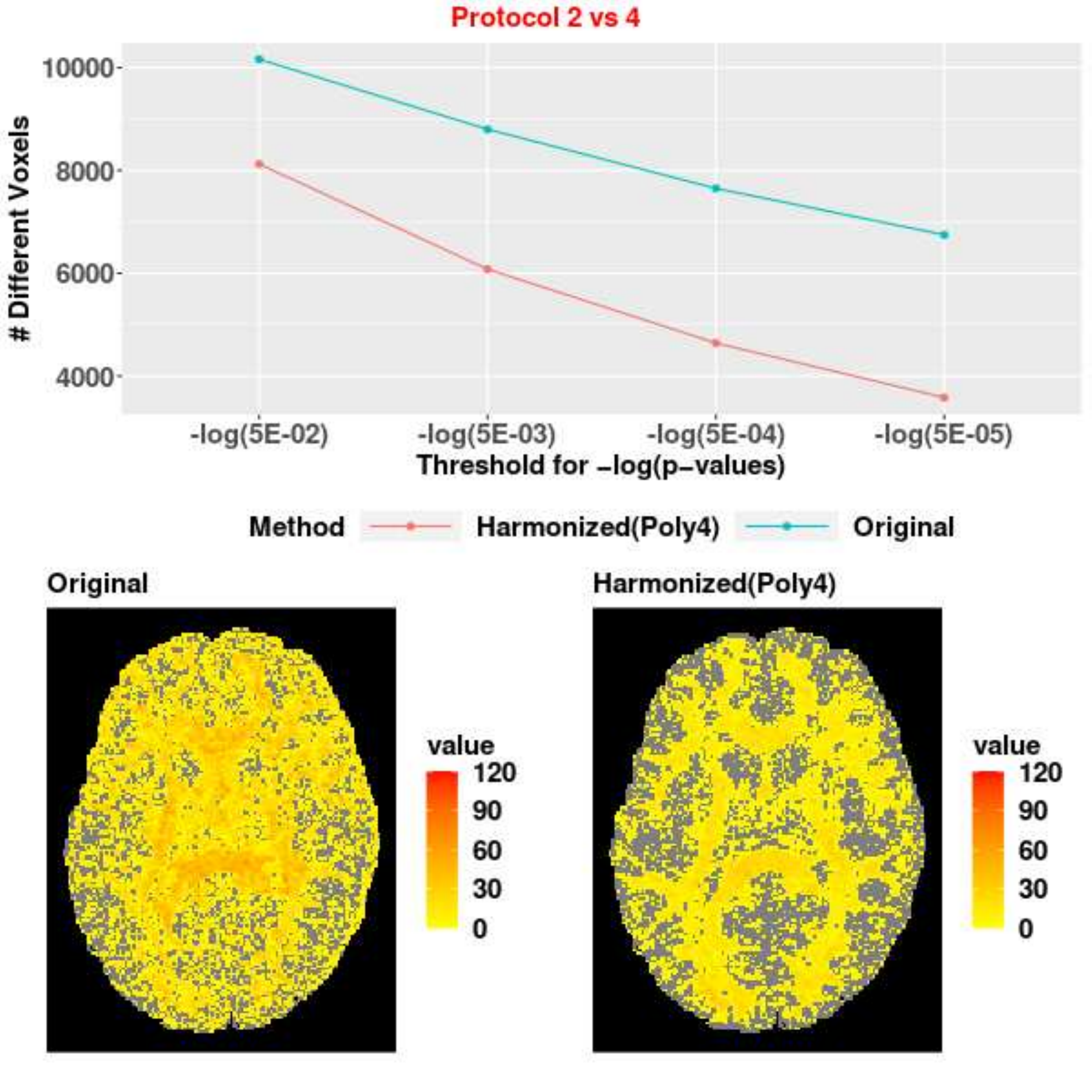}
    \caption{Evaluation of difference of NODDI metrics produced by protocols 2 and 4. The first figure on the first row shows the number of voxels with $-log(\text{p-value})$ of two-sample paired t-tests more than a given threshold.
    The other pair of images on the second row present voxel-wise $-log(\text{p-value})$ of two-sample paired t-tests. The harmonized data is obtained by applying our proposed model with $K=4$.
    }
    \label{fig:harmo1}
\end{figure}

\begin{figure}
    \centering
    \includegraphics[width = 0.9\textwidth]{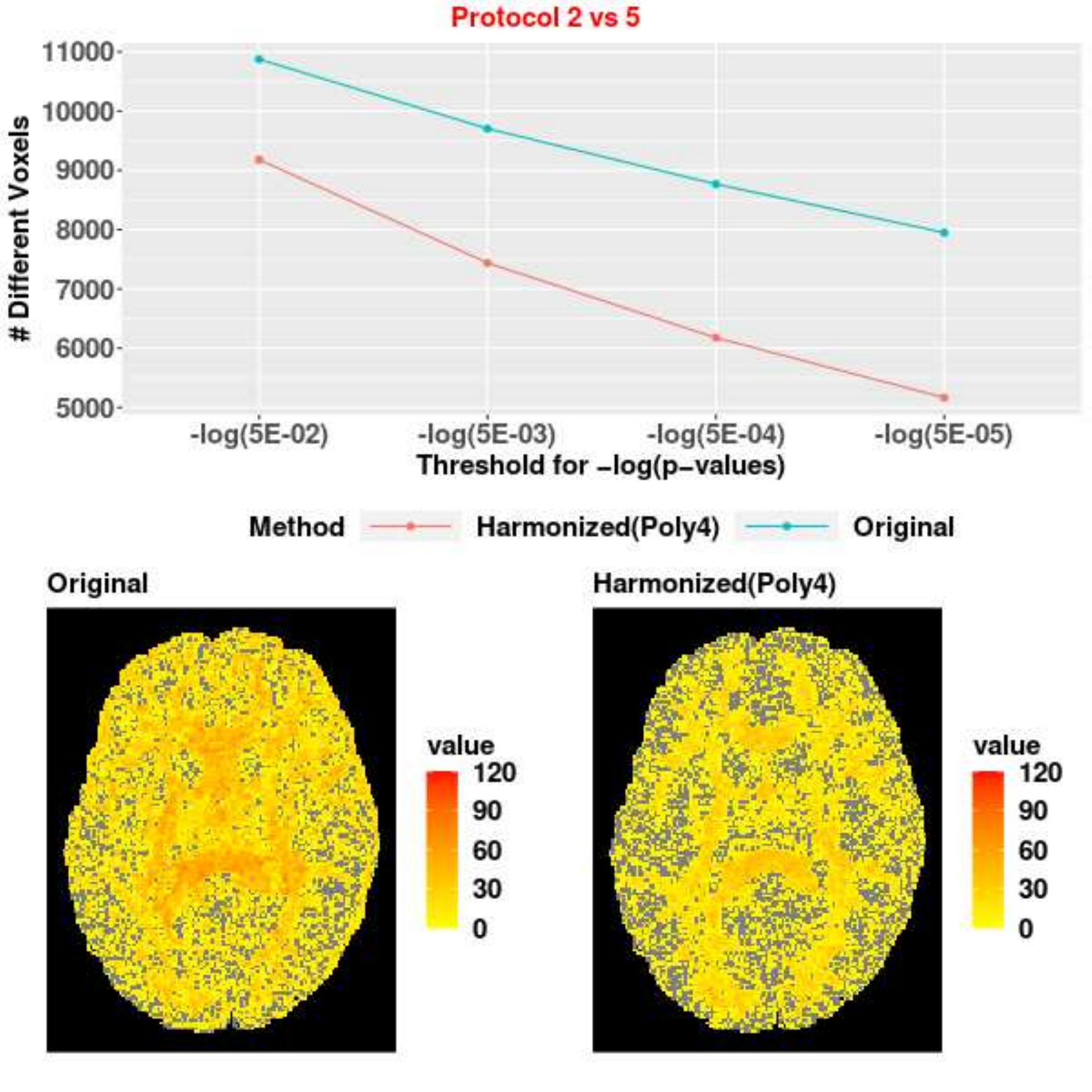}
    \caption{Evaluation of difference of NODDI metrics produced by protocols 2 and 5. The first figure on the first row shows the number of voxels with $-log(\text{p-value})$ of two-sample paired t-tests more than a given threshold.
    The other pair of images on the second row present voxel-wise $-log(\text{p-value})$ of two-sample paired t-tests. The harmonized data is obtained by applying our proposed model with $K=4$.
    }
    \label{fig:harmo1}
\end{figure}

\begin{figure}
    \centering
    \includegraphics[width = 0.9\textwidth]{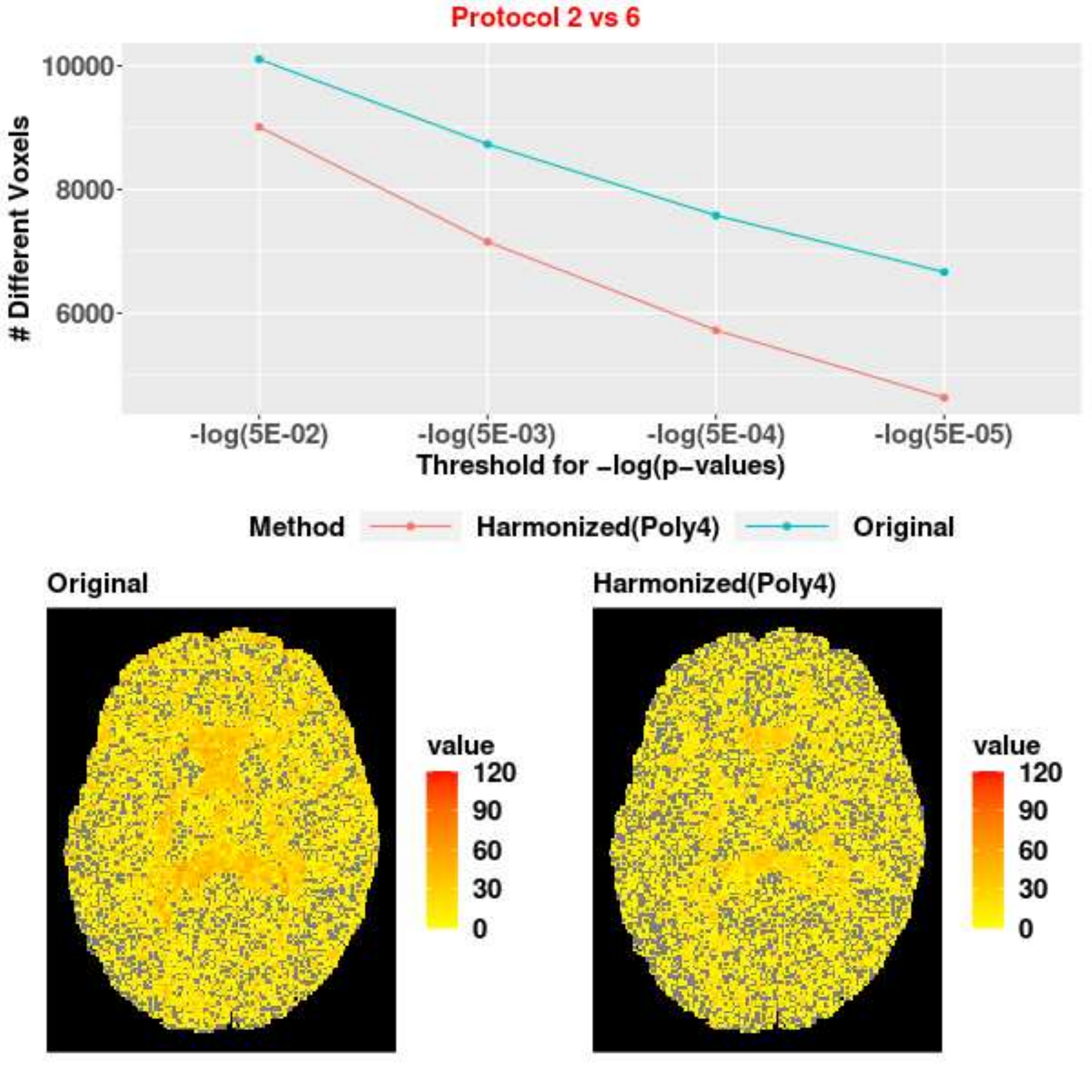}
    \caption{Evaluation of difference of NODDI metrics produced by protocols 2 and 6. The first figure on the first row shows the number of voxels with $-log(\text{p-value})$ of two-sample paired t-tests more than a given threshold.
    The other pair of images on the second row present voxel-wise $-log(\text{p-value})$ of two-sample paired t-tests. The harmonized data is obtained by applying our proposed model with $K=4$.
    }
    \label{fig:harmo1}
\end{figure}

\begin{figure}
    \centering
    \includegraphics[width = 0.9\textwidth]{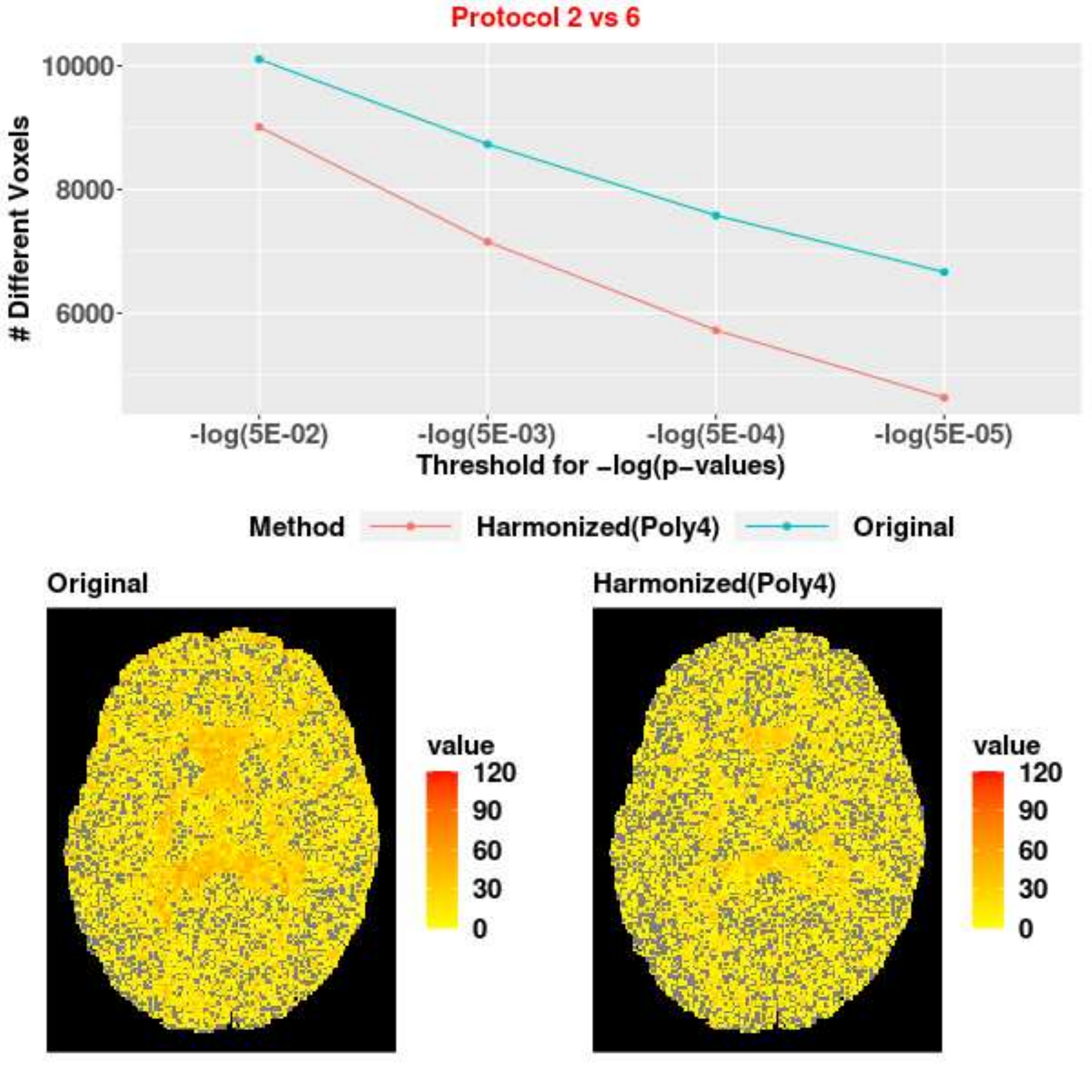}
    \caption{Evaluation of difference of NODDI metrics produced by protocols 2 and 6. The first figure on the first row shows the number of voxels with $-log(\text{p-value})$ of two-sample paired t-tests more than a given threshold.
    The other pair of images on the second row present voxel-wise $-log(\text{p-value})$ of two-sample paired t-tests. The harmonized data is obtained by applying our proposed model with $K=4$.
    }
    \label{fig:harmo1}
\end{figure}

\begin{figure}
    \centering
    \includegraphics[width = 0.9\textwidth]{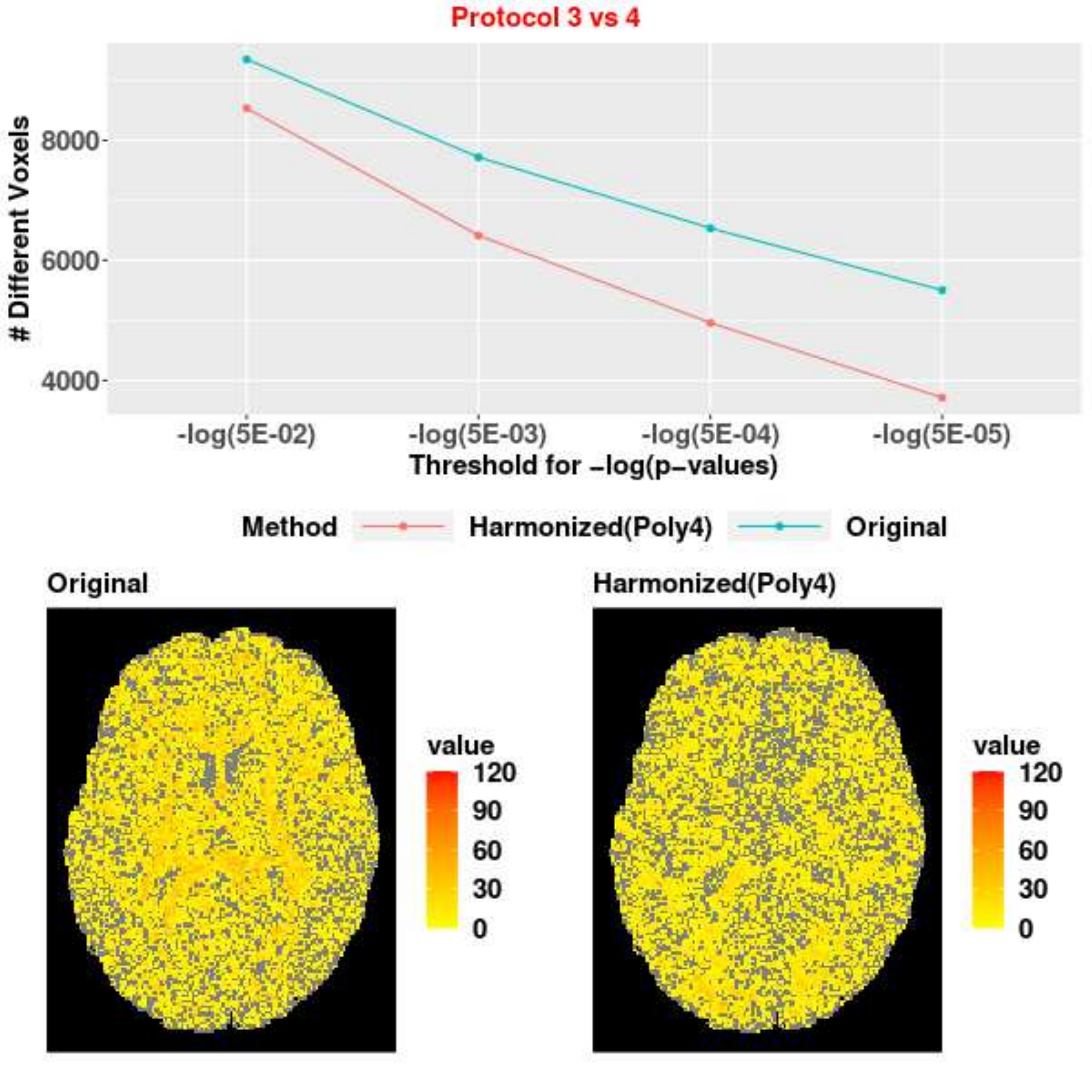}
    \caption{Evaluation of difference of NODDI metrics produced by protocols 3 and 4. The first figure on the first row shows the number of voxels with $-log(\text{p-value})$ of two-sample paired t-tests more than a given threshold.
    The other pair of images on the second row present voxel-wise $-log(\text{p-value})$ of two-sample paired t-tests. The harmonized data is obtained by applying our proposed model with $K=4$.
    }
    \label{fig:harmo1}
\end{figure}

\begin{figure}
    \centering
    \includegraphics[width = 0.9\textwidth]{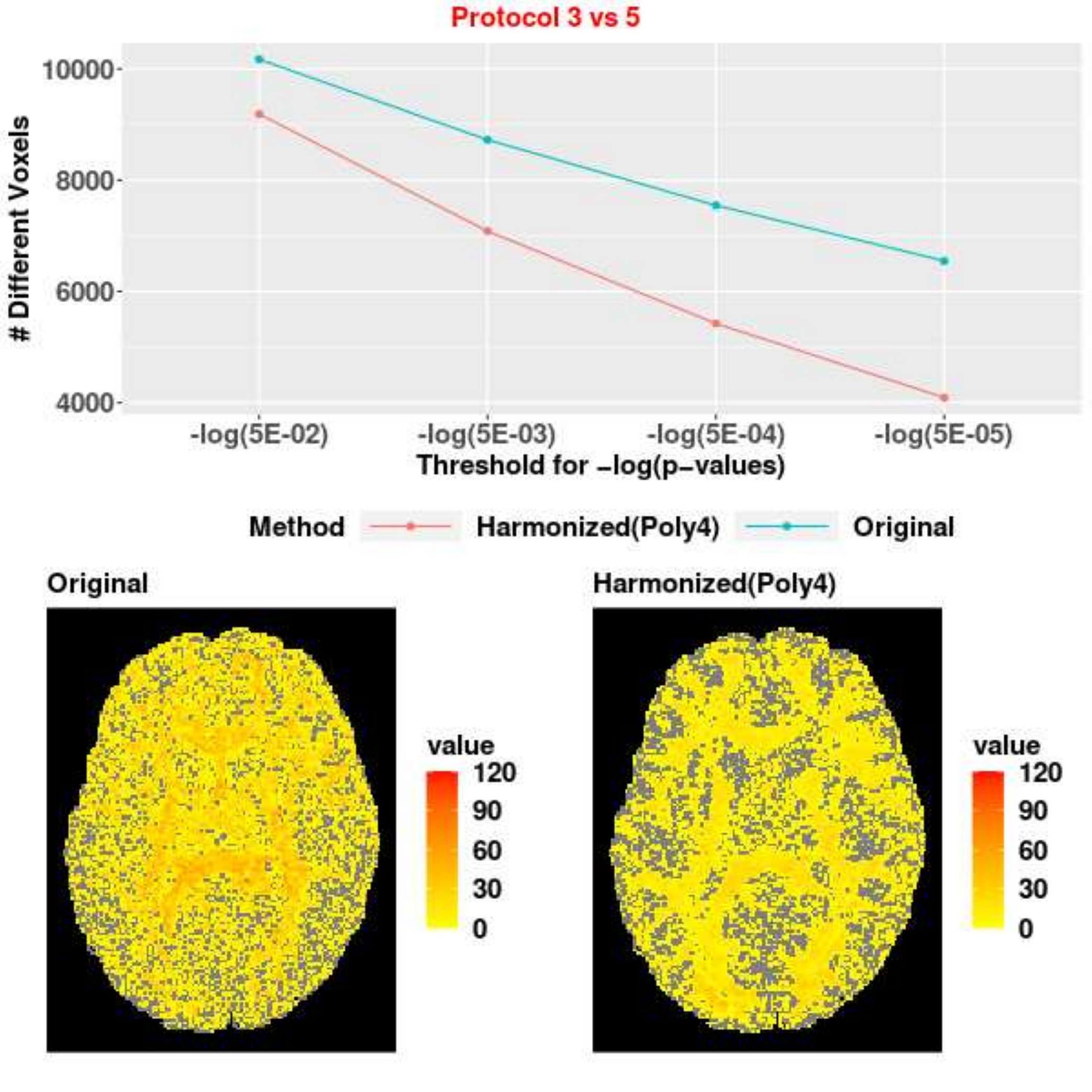}
    \caption{Evaluation of difference of NODDI metrics produced by protocols 3 and 5. The first figure on the first row shows the number of voxels with $-log(\text{p-value})$ of two-sample paired t-tests more than a given threshold.
    The other pair of images on the second row present voxel-wise $-log(\text{p-value})$ of two-sample paired t-tests. The harmonized data is obtained by applying our proposed model with $K=4$.
    }
    \label{fig:harmo1}
\end{figure}

\begin{figure}
    \centering
    \includegraphics[width = 0.9\textwidth]{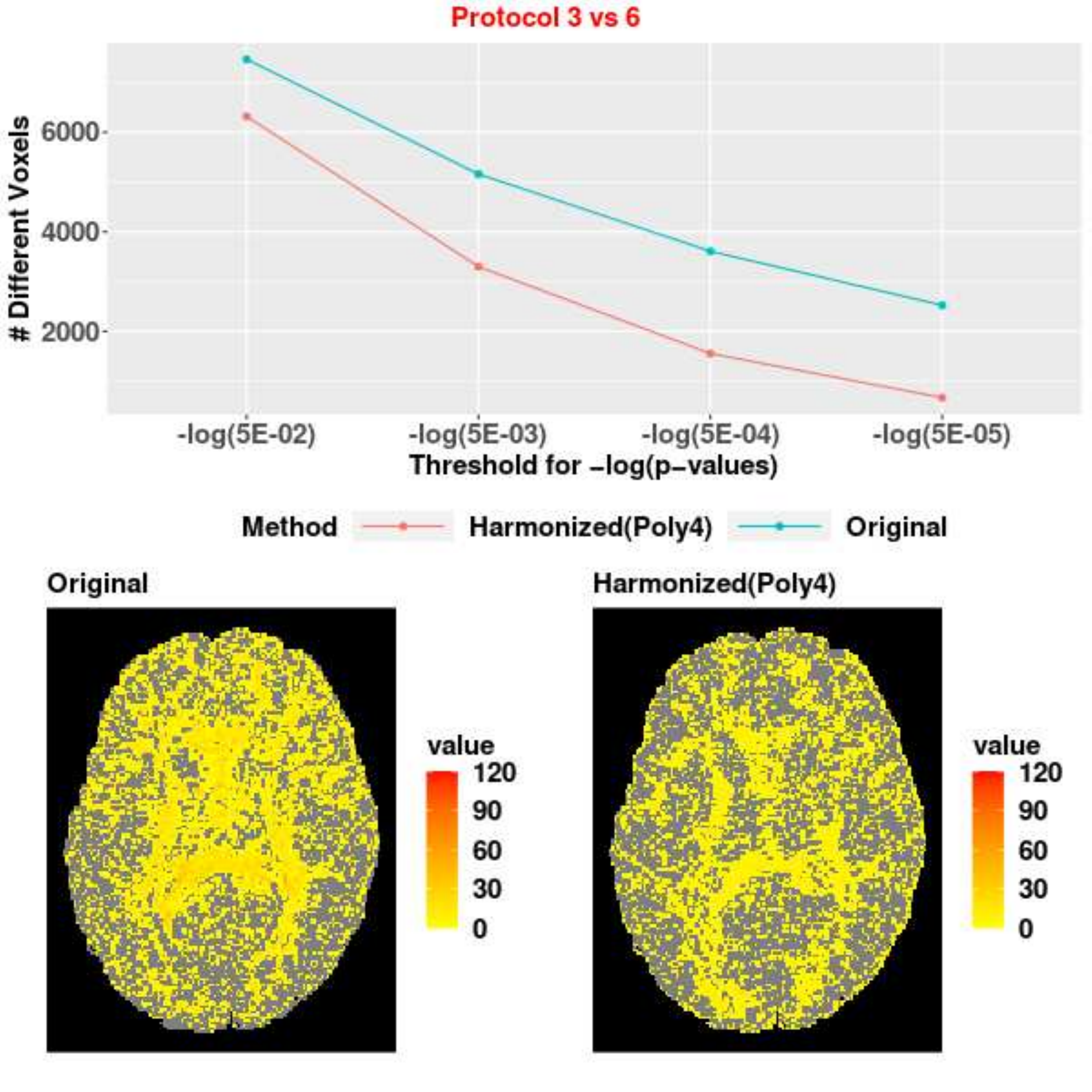}
    \caption{Evaluation of difference of NODDI metrics produced by protocols 3 and 6. The first figure on the first row shows the number of voxels with $-log(\text{p-value})$ of two-sample paired t-tests more than a given threshold.
    The other pair of images on the second row present voxel-wise $-log(\text{p-value})$ of two-sample paired t-tests. The harmonized data is obtained by applying our proposed model with $K=4$.
    }
    \label{fig:harmo1}
\end{figure}

\begin{figure}
    \centering
    \includegraphics[width = 0.9\textwidth]{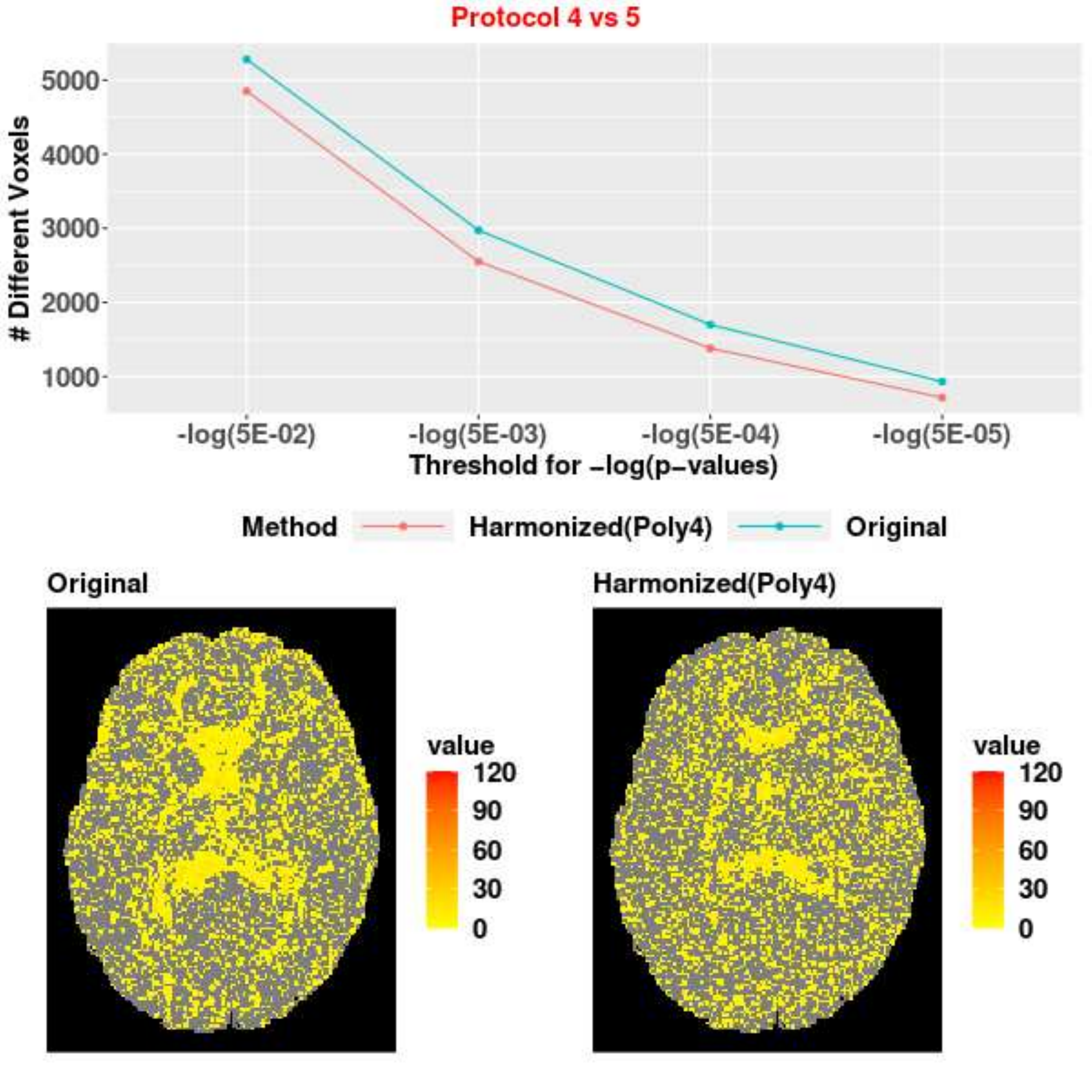}
    \caption{Evaluation of difference of NODDI metrics produced by protocols 4 and 6. The first figure on the first row shows the number of voxels with $-log(\text{p-value})$ of two-sample paired t-tests more than a given threshold.
    The other pair of images on the second row present voxel-wise $-log(\text{p-value})$ of two-sample paired t-tests. The harmonized data is obtained by applying our proposed model with $K=4$.
    }
    \label{fig:harmo1}
\end{figure}

\begin{figure}
    \centering
    \includegraphics[width = 0.9\textwidth]{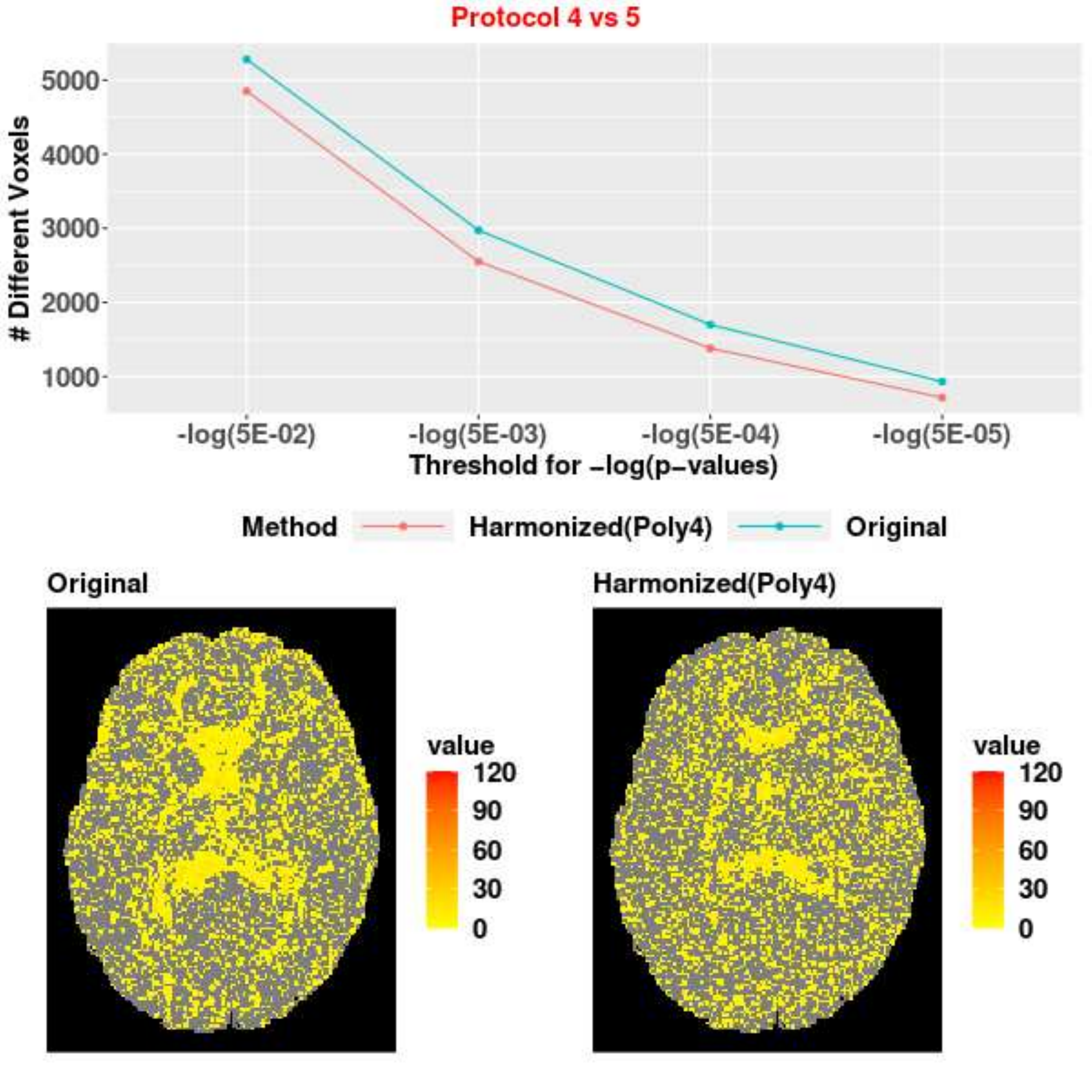}
    \caption{Evaluation of difference of NODDI metrics produced by protocols 4 and 5. The first figure on the first row shows the number of voxels with $-log(\text{p-value})$ of two-sample paired t-tests more than a given threshold.
    The other pair of images on the second row present voxel-wise $-log(\text{p-value})$ of two-sample paired t-tests. The harmonized data is obtained by applying our proposed model with $K=4$.
    }
    \label{fig:harmo1}
\end{figure}

\begin{figure}
    \centering
    \includegraphics[width = 0.9\textwidth]{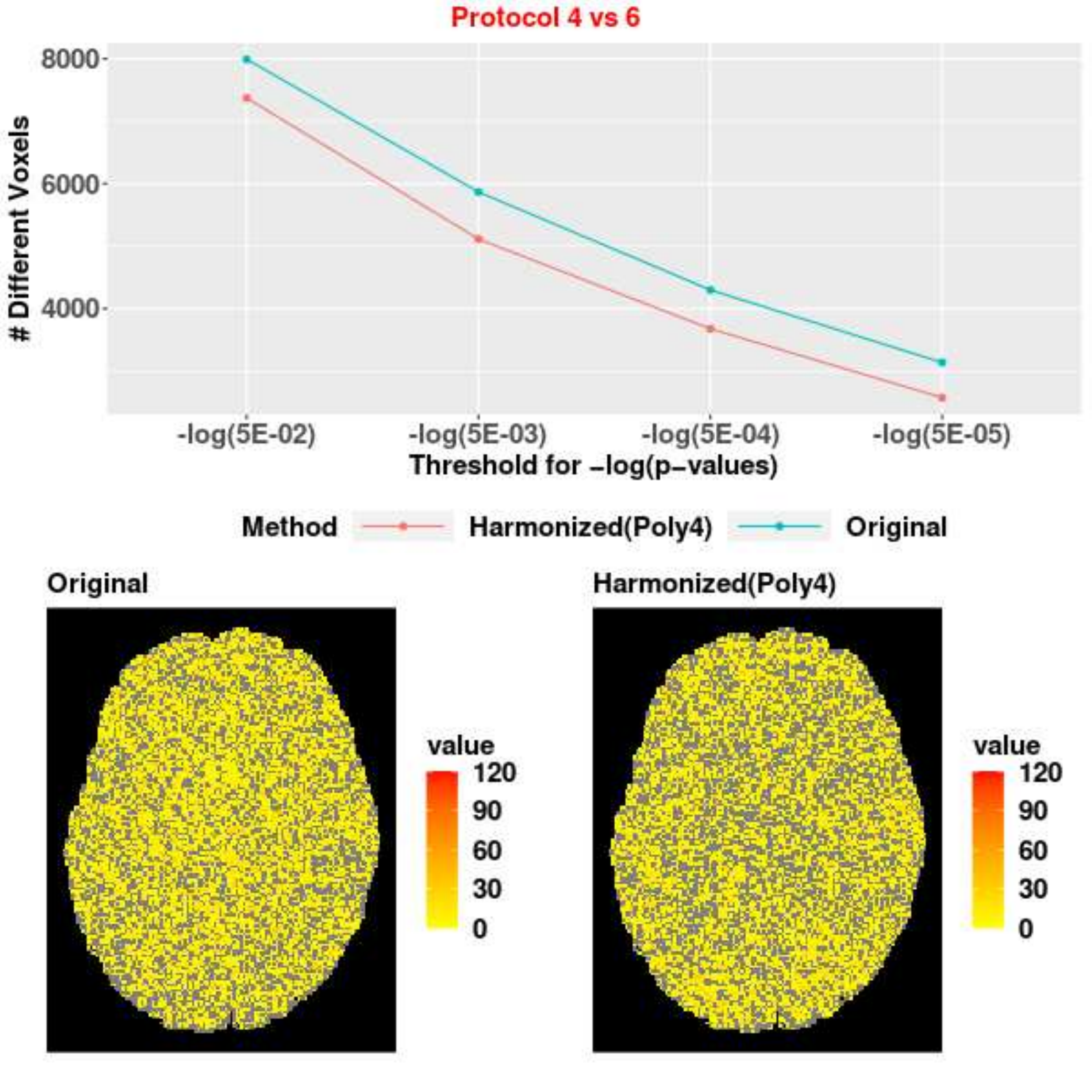}
    \caption{Evaluation of difference of NODDI metrics produced by protocols 4 and 6. The first figure on the first row shows the number of voxels with $-log(\text{p-value})$ of two-sample paired t-tests more than a given threshold.
    The other pair of images on the second row present voxel-wise $-log(\text{p-value})$ of two-sample paired t-tests. The harmonized data is obtained by applying our proposed model with $K=4$.
    }
    \label{fig:harmo1}
\end{figure}

\begin{figure}
    \centering
    \includegraphics[width = 0.9\textwidth]{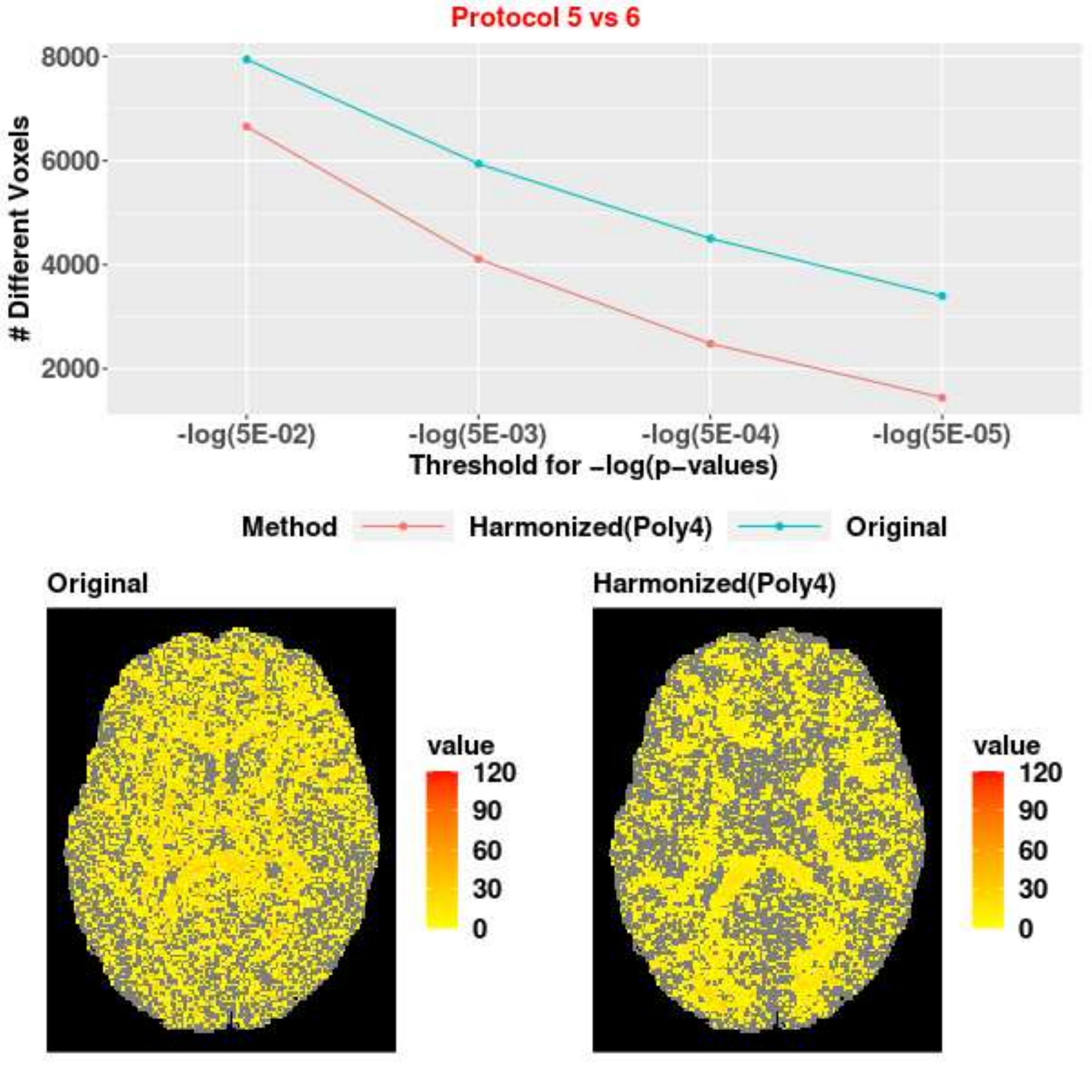}
    \caption{Evaluation of difference of NODDI metrics produced by protocols 5 and 6. The first figure on the first row shows the number of voxels with $-log(\text{p-value})$ of two-sample paired t-tests more than a given threshold.
    The other pair of images on the second row present voxel-wise $-log(\text{p-value})$ of two-sample paired t-tests. The harmonized data is obtained by applying our proposed model with $K=4$.
    }
    \label{fig:harmo1}
\end{figure}

\begin{figure}
    \centering
    \includegraphics[width = 1\textwidth]{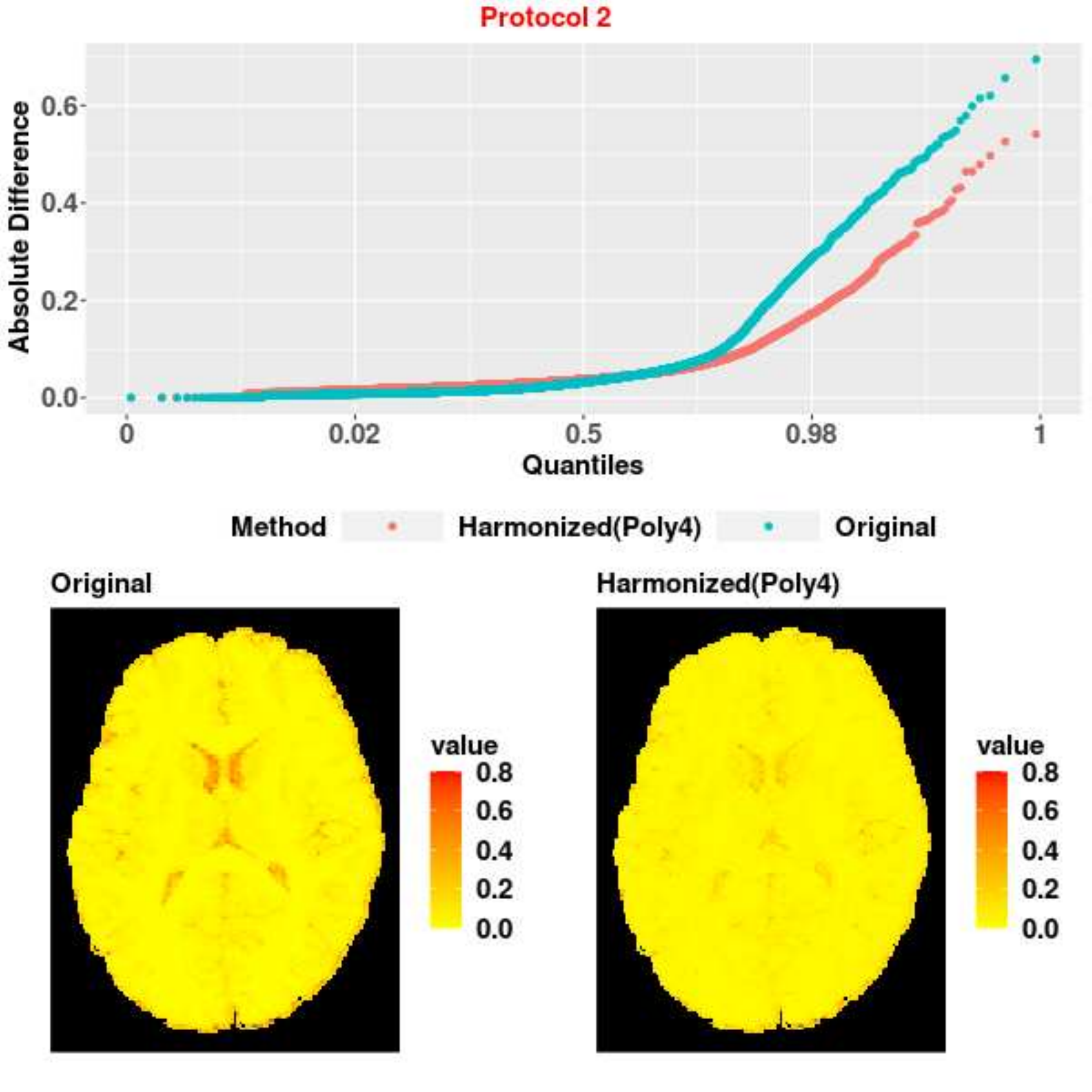}
    \caption{Protocol 2: 
    The figure in the first row shows the quantiles of the mean absolute difference of ODI estimates between the gold standard one and those obtained from the original and harmonized re-sampled dMRI data. The two images on the second row are the maps of the absolute difference based on the original signals and harmonized signals, respectively, for comparison. The harmonized data is obtained by applying our proposed method with $K=4$.
    }
    \label{fig:absdiff1}
\end{figure}

\begin{figure}
    \centering
    \includegraphics[width = 1\textwidth]{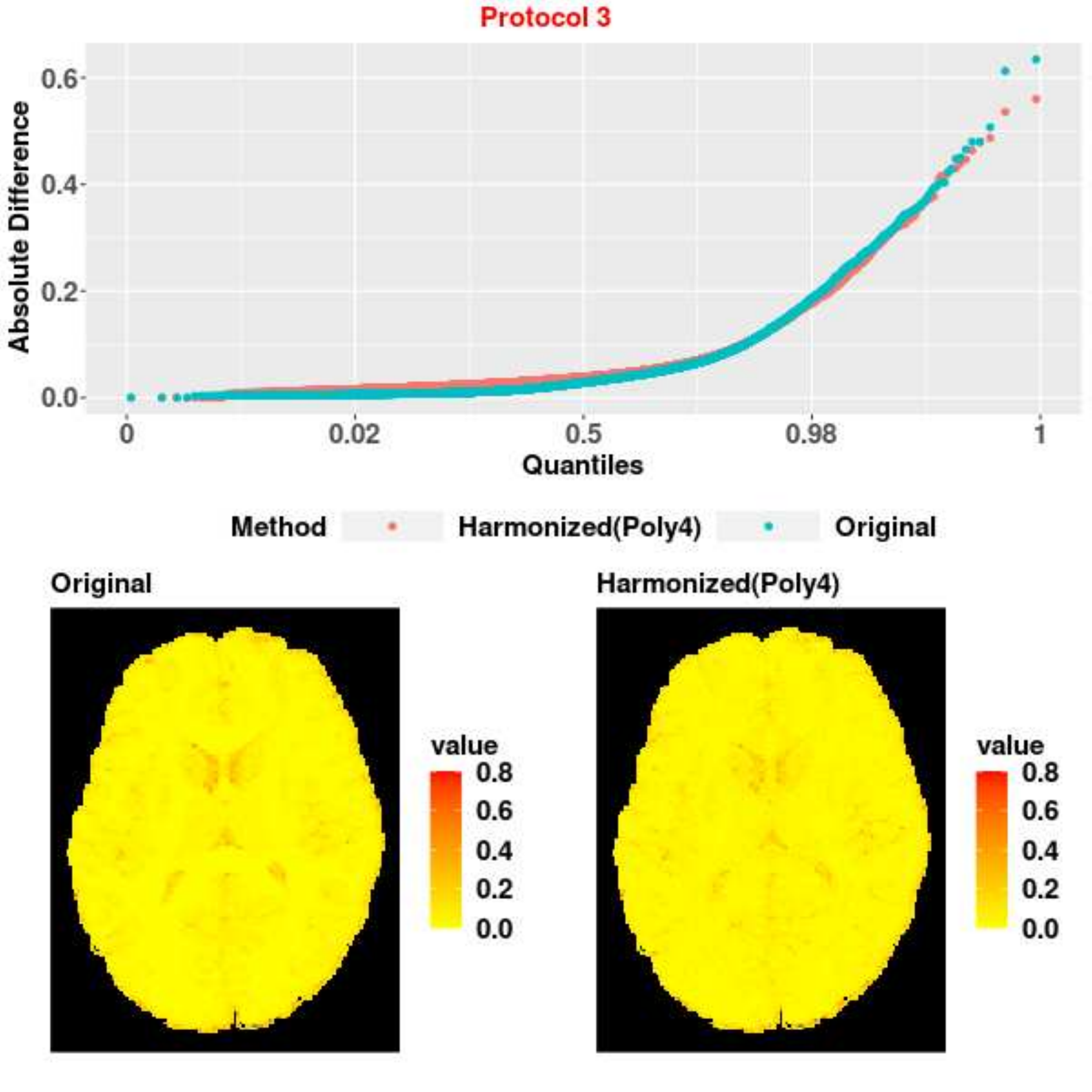}
    \caption{Protocol 3: 
    The figure in the first row shows the quantiles of the mean absolute difference of ODI estimates between the gold standard one and those obtained from the original and harmonized re-sampled dMRI data. The two images on the second row are the maps of the absolute difference based on the original signals and harmonized signals, respectively, for comparison. The harmonized data is obtained by applying our proposed method with $K=4$.
    }
    \label{fig:absdiff1}
\end{figure}

\begin{figure}
    \centering
    \includegraphics[width = 1\textwidth]{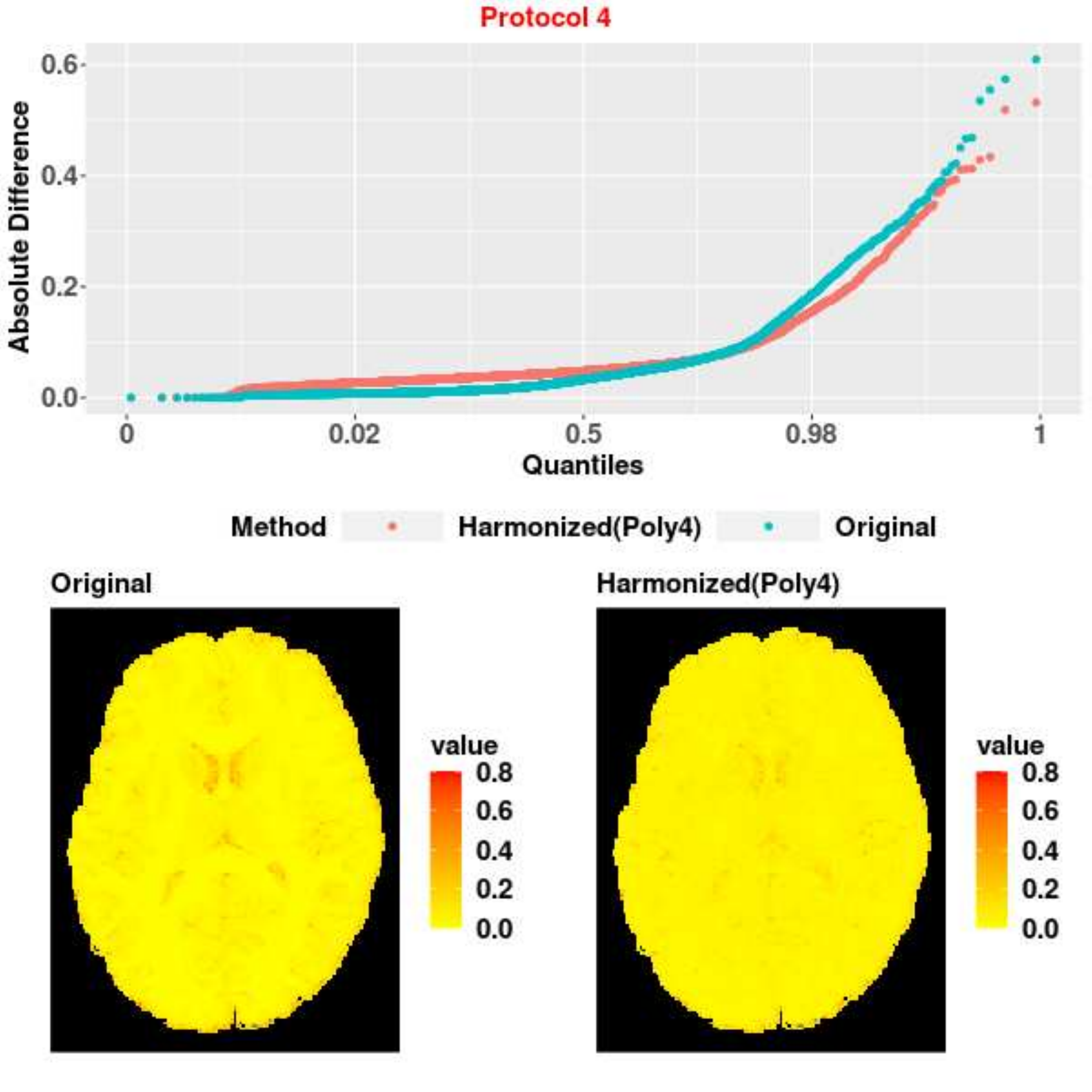}
    \caption{Protocol 4: 
    The figure in the first row shows the quantiles of the mean absolute difference of ODI estimates between the gold standard one and those obtained from the original and harmonized re-sampled dMRI data. The two images on the second row are the maps of the absolute difference based on the original signals and harmonized signals, respectively, for comparison. The harmonized data is obtained by applying our proposed method with $K=4$.
    }
    \label{fig:absdiff1}
\end{figure}

\begin{figure}
    \centering
    \includegraphics[width = 1\textwidth]{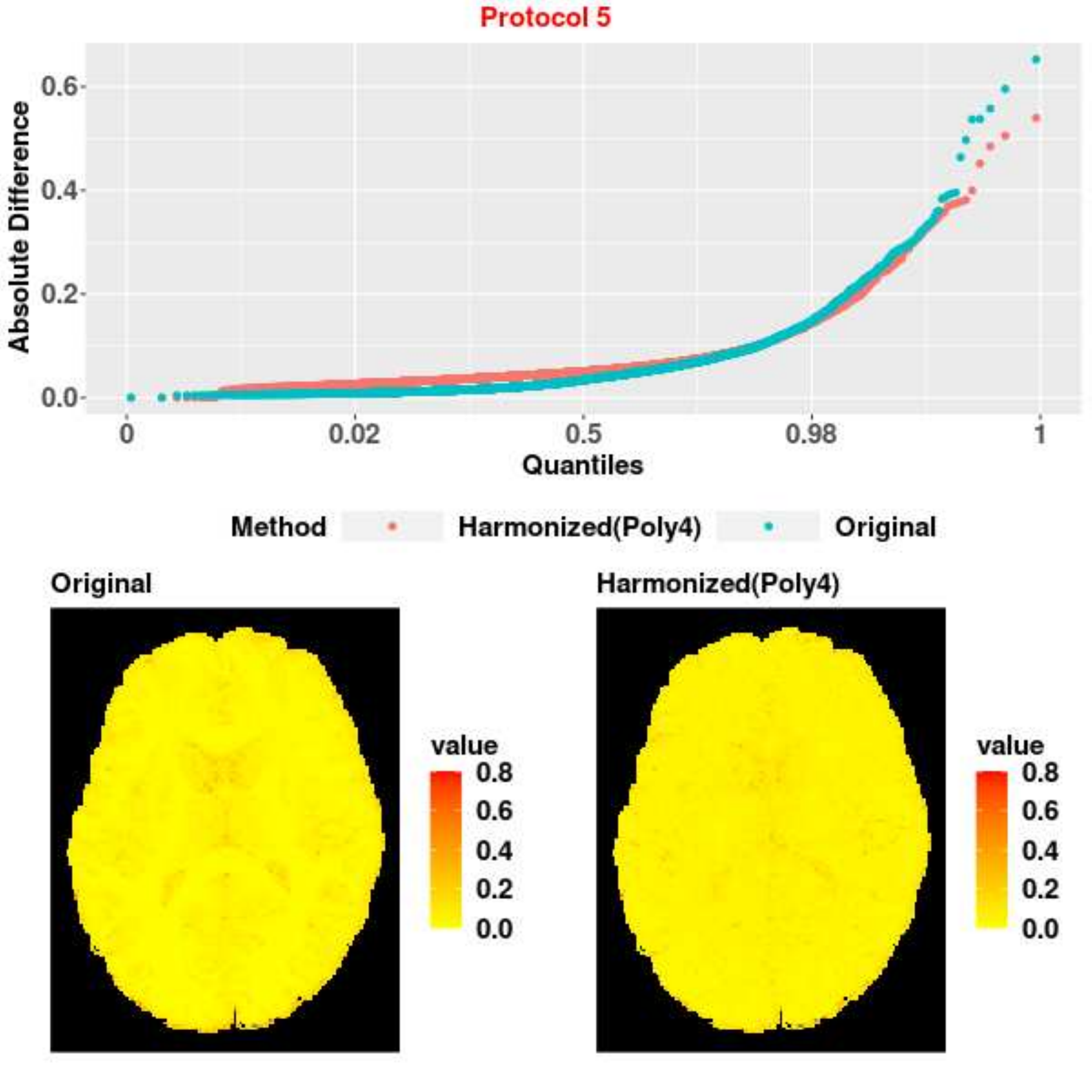}
    \caption{Protocol 5: 
    The figure in the first row shows the quantiles of the mean absolute difference of ODI estimates between the gold standard one and those obtained from the original and harmonized re-sampled dMRI data. The two images on the second row are the maps of the absolute difference based on the original signals and harmonized signals, respectively, for comparison. The harmonized data is obtained by applying our proposed method with $K=4$.
    }
    \label{fig:absdiff1}
\end{figure}

\begin{figure}
    \centering
    \includegraphics[width = 1\textwidth]{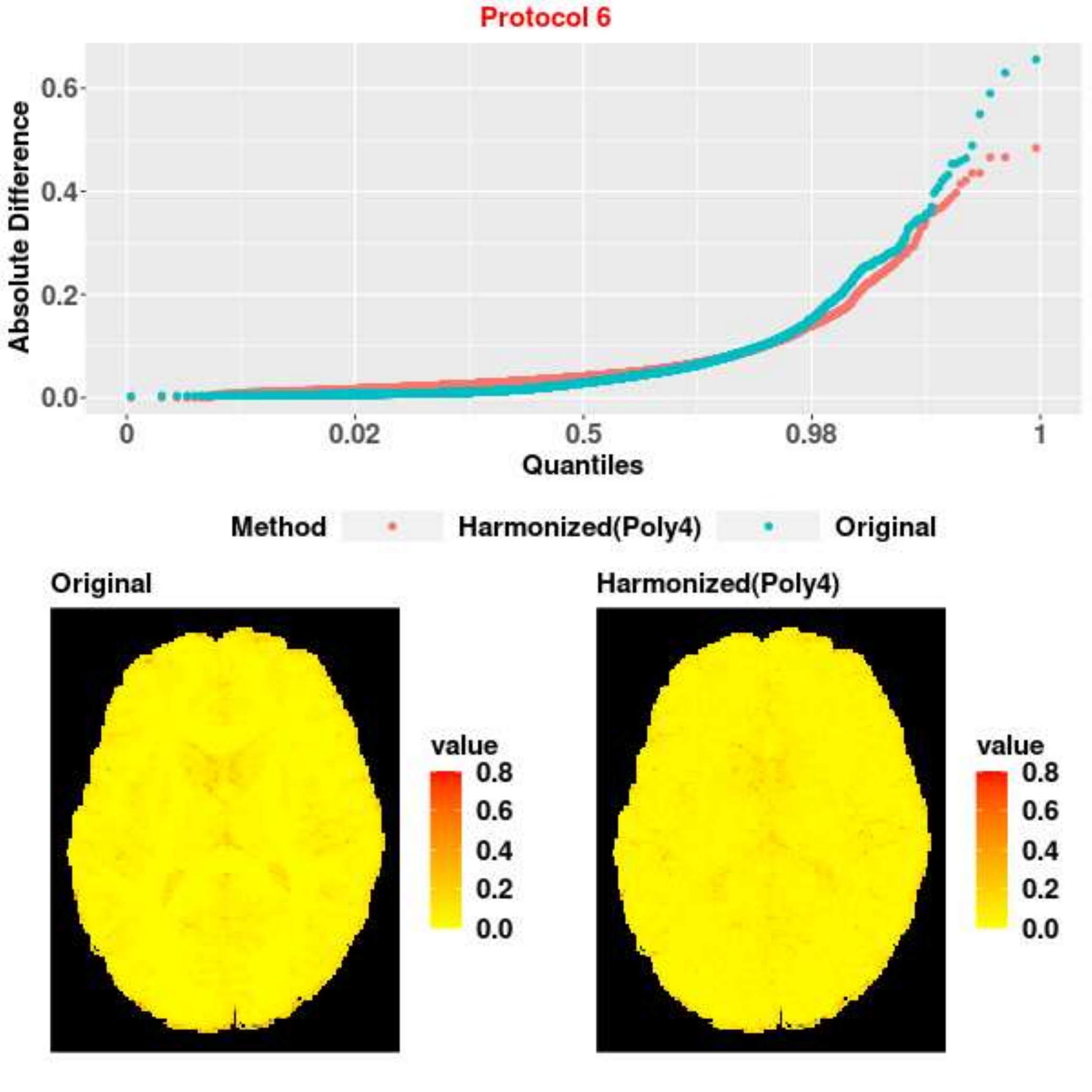}
    \caption{Protocol 6: 
    The figure in the first row shows the quantiles of the mean absolute difference of ODI estimates between the gold standard one and those obtained from the original and harmonized re-sampled dMRI data. The two images on the second row are the maps of the absolute difference based on the original signals and harmonized signals, respectively, for comparison. The harmonized data is obtained by applying our proposed method with $K=4$.
    }
    \label{fig:absdiff1}
\end{figure}






\bibliography{sbci_bib,paper,paperdti}




\end{document}